\documentclass[11pt]{article}

\usepackage{graphicx}
\usepackage{latexsym}
\usepackage{amsmath}
\usepackage{amssymb}

\usepackage{url}
\usepackage{amsthm}

\newcommand{\Var}{\mbox{Var}}

\newcommand{\bA}{\mbox{\boldmath {$A$}}}

\newcommand{\bB}{\mbox{\boldmath {$B$}}}

\newcommand{\bE}{\mbox{\boldmath {$E$}}}

\newcommand{\bh}{\mbox{\boldmath {$h$}}}
\newcommand{\bH}{\mbox{\boldmath {$H$}}}

\newcommand{\bI}{\mbox{\boldmath {$I$}}}

\newcommand{\bO}{\mbox{\boldmath {$O$}}}

\newcommand{\bP}{\mbox{\boldmath {$P$}}}

\newcommand{\bThe}{\mbox{\boldmath $\Theta$}}
\newcommand{\bsR}{\mbox{\bf R}}

\newcommand{\bS}{\mbox{\boldmath {$S$}}}

\newcommand{\bV}{\mbox{\boldmath {$V$}}}
\newcommand{\bw}{\mbox{\boldmath {$w$}}}

\newcommand{\bx}{\mbox{\boldmath {$x$}}}
\newcommand{\bX}{\mbox{\boldmath {$X$}}}

\newcommand{\bY}{\mbox{\boldmath {$Y$}}}
\newcommand{\bz}{\mbox{\boldmath {$z$}}}

\newcommand{\bze}{\mbox{\boldmath {$0$}}}

\newcommand{\bmu}{\mbox{\boldmath $ \mu $}}
\newcommand{\bSig}{\mbox{\boldmath $ \Sigma $}}

\newcommand{\bSigma}{\mbox{\boldmath $ \Sigma $}}

\newcommand{\bLam}{\mbox{\boldmath $ \Lambda $}}

\newcommand{\bGamma}{\mbox{\boldmath $\Gamma$}}

\newcommand{\bgamma}{\mbox{\boldmath $\gamma$}}

\newcommand{\bem   }{\mbox{\boldmath $ \emptyset  $}}
\newcommand{\tr}{\mbox{tr}}

\theoremstyle{plain}

\newtheorem{thm}{Theorem}[section]
\newtheorem{cor}{Corollary}[section]
\newtheorem{lem}{Lemma}[section]
\newtheorem{pro}{Proposition}[section]

\long\def\symbolfootnote[#1]#2{\begingroup%
\def\thefootnote{\fnsymbol{footnote}}\footnote[#1]{#2}\endgroup}

\begin{document}

\begin{center}
\Large
{\bf High-dimensional inference on covariance structures\\ via the extended cross-data-matrix methodology}
\end{center}
%%%%%%%%% Authors, affiliations %%%%%%%%%%%%%%%%%%%%%%%%%%
\begin{center}
\vskip 0.5cm
\textbf{\large Kazuyoshi Yata and Makoto Aoshima} \\
Institute of Mathematics, University of Tsukuba, Ibaraki, Japan \\[-1cm]
\end{center}
\symbolfootnote[0]{\normalsize Address correspondence to Makoto Aoshima, 
Institute of Mathematics, University of Tsukuba, Ibaraki 305-8571, Japan; 
Fax: +81-298-53-6501; E-mail: aoshima@math.tsukuba.ac.jp}


\begin{abstract}
In this paper, we consider testing the correlation coefficient matrix between two subsets of high-dimensional variables.
We produce a test statistic by using the extended cross-data-matrix (ECDM) methodology and show the unbiasedness of ECDM estimator.
We also show that the ECDM estimator has the consistency property and the asymptotic normality in high-dimensional settings.  
We propose a test procedure by the ECDM estimator and evaluate its asymptotic size and power theoretically and numerically. 
We give several applications of the ECDM estimator.
Finally, we demonstrate how the test procedure performs in actual data analyses by using a microarray data set. \\
\\
{\small \noindent\textbf{Keywords:} Correlations test; Cross-data-matrix methodology; Graphical modeling; Large $p$, small $n$; Pathway analysis;  RV-coefficient}
\end{abstract}

\section{Introduction}
\label{1}
%%%%%%%%%%%%%%%%%%%%%%%
Suppose we take samples, $\bx_{j},\ j=1,...,n$, of size $n\ (\ge 4)$, which are independent and identically distributed (i.i.d.) as a $p$-variate distribution. 
Here, we consider situations where the data dimension $p$ is very high compared to the sample size $n$. 
Let $\bx_{j}=(\bx_{1j}^T,\bx_{2j}^T)^T$ and assume $\bx_{ij}\in \bsR^{p_i}$, $i=1,2$, with  $p_1\in [1,p-1]$ and $p_2=p-p_1$. 
We assume that $\bx_j$ has an unknown mean vector, $\bmu=(\bmu_1^T,\bmu_2^T)^T$, and unknown covariance matrix, 
$$
\bSig=\begin{pmatrix} \bSig_1& \bSig_{*} \\ \bSig_{*}^T & \bSig_2 \end{pmatrix}\ (\ge \bO),
$$ 
that is, $E(\bx_{ij})=\bmu_i$, $\Var(\bx_{ij})=\bSig_i$, $i=1,2,$ and Cov$(\bx_{1j},\bx_{2j})=E(\bx_{1j} \bx_{2j}^T)-\bmu_1\bmu_2^T=\bSig_{*}$. 
Let $\sigma_{ij}$ be the $j$-th diagonal element of $\bSig_i$ for $i=1,2;\ j=1,...,p_i$, and assume $\sigma_{ij}>0$ for all $i,j$. 
We denote the correlation coefficient matrix between $\bx_{1j}$ and $\bx_{2j}$ by $\mbox{Corr}(\bx_{1j},\bx_{2j})=\bP$, where $\bP=\mbox{diag}(\sigma_{11},...,\sigma_{1p_1})^{-1/2}\bSig_{*}$
$\mbox{diag}(\sigma_{21},...,\sigma_{2p_2})^{-1/2}$. 
Here, $\mbox{diag}(\sigma_{i1},...,\sigma_{ip_i})$ denotes the diagonal matrix of elements, $\sigma_{i1},...,\sigma_{ip_i}$.

In this paper, we consider testing the correlation coefficient matrix between $\bx_{1j}$ and $\bx_{2j}$ by 
\begin{align}
H_0: \bP=\bO \quad \mbox{vs.}\quad H_1:\bP \neq \bO \label{1.1}
\end{align}
for high-dimensional settings. 
When $(p_1,p_2)=(p-1,1)$ or $(1,p-1)$, (\ref{1.1}) implies the test of correlation coefficients. 
Aoshima and Yata \cite{Aoshima:2011} gave a test statistic for the test of correlation coefficients and Yata and Aoshima \cite{Yata:2013} improved the test statistic by using a method called the {\it extended cross-data-matrix (ECDM) methodology}.
The test of correlation coefficient matrix is a very important tool of pathway analysis or graphical modeling for high-dimensional data. 
One of the applications is to construct gene networks. 
See Figure 1. 
\begin{figure}
%%%%%%%%%%%%%%%%%%%%%%%%%%%%%%%%%%
\includegraphics[scale=0.5]{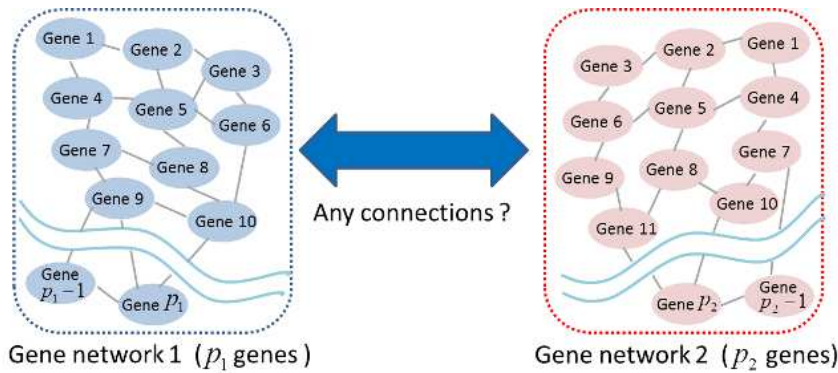}
\caption{
Illustration of the test by (\ref{1.1}). 
On can apply the test to constructing gene networks.} 
\end{figure}
%%%%%%%%%%%%%%%%%%%%%%%%%%%%%%%%%%
Drton and Perlman \cite{Drton:2007} and Wille et al. \cite{Wille:2004} considered pathway analysis or graphical modeling of microarray data by testing an individual correlation coefficient. 
For example, Wille et al. \cite{Wille:2004} analyzed gene networks of microarray data with $p=834$ ($p_1=39$ and $p_2=795$) and $n=118$. 
On the other hand, Hero and Rajaratnam \cite{Hero:2011} considered correlation screening procedures for high-dimensional data by using a test of correlations. 
Lan et al. \cite{Lan:2014} and Zhong and Chen \cite{Zhong:2011} considered tests of regression coefficient vectors in linear regression models. 
As for the test of independence, see Fujikoshi et al. \cite{Fujikoshi:2010}, 
Srivastava and Reid \cite{Srivastava:2012} and Yang and Pan \cite{Yang:2015}. 
Also, one may refer to Sz\'{e}kely and Rizzo \cite{Szekely:2009,Szekely:2013} about distance correlation. 

In Section 2, we give several assumptions to construct a high-dimensional correlation test for (\ref{1.1}). 
In Section 3, we produce a test statistic for (\ref{1.1}) by using the ECDM methodology and show the unbiasedness of ECDM estimator.
We also show that the ECDM estimator has the consistency property and the asymptotic normality when $p\to\infty$ and $n\to\infty$. 
In Section 4, we propose a test procedure for (\ref{1.1}) by the ECDM estimator and evaluate its asymptotic size and power when $p\to\infty$ and $n\to\infty$ theoretically and numerically. 
In Section 5, we give several applications of the ECDM estimator.
Finally, we demonstrate how the test procedure performs in actual data analyses by using a microarray data set. 
%%%%%%%%%%%%%%%%%%%%%%
\section{Assumptions}
%%%%%%%%%%%%%%%%%%%%%%
In this section, we give several assumptions to construct a test procedure for (\ref{1.1}).  
We have the eigenvalue decomposition of $\bSig$ by $\bSig=\bH\bLam \bH^T$, where $\bLam=$diag$(\lambda_{1},...,\lambda_{p})$ having eigenvalues, $\lambda_{1}\ge \cdots \ge \lambda_{p}\ge 0$, and $\bH$ is an orthogonal matrix of the corresponding eigenvectors. 
Let $\bx_j=\bH \bLam^{1/2}\bz_j+\bmu$, $j=1,...,n$, where $E(\bz_{j})=\bze$ and $\Var(\bz_{j})=\bI_{p}$.
Here, $\bI_p$ denotes the identity matrix of dimension $p$. 
Note that if $\bx_j$ is Gaussian, the elements of $\bz_{j}$ are i.i.d. as the standard normal distribution, $N(0,1)$. 
We assume the following model: 
\begin{equation}
\bx_j=\bGamma \bw_{j}+\bmu, \ j=1,...,n,
\label{2.1}
\end{equation}
where $\bGamma$ is a $p\times q$ matrix for some $q>0$ such that $\bGamma \bGamma^T=\bSigma$, and $\bw_{j}=(w_{1j},...,w_{qj})^T,\ j=1,...,n$, are i.i.d. random vectors having $E(\bw_{j})=\bze$ and $\Var(\bw_{j})=\bI_{q}$. 
Let $\bGamma=(\bGamma_1^T,\bGamma_2^T)^T$, where $\bGamma_i=(\bgamma_{i1},...,\bgamma_{iq})$ with $\bgamma_{ij}$s $\in \bsR^{p_i}$ for $i=1,2$.
Then, we have that $\bx_{ij}=\bGamma_i \bw_{j}+\bmu_i$ for $i=1,2$. 
Note that $\bSig_{*}=\bGamma_1 \bGamma_2^T=\sum_{r=1}^q \bgamma_{1r}\bgamma_{2r}^T$. 
Also, note that (\ref{2.1}) includes the case that $\bGamma=\bH \bLam^{1/2}$ and $\bw_{j}=\bz_{j}$. 
Let $\Var(w_{rj}^2)=M_r$, $r=1,...,q$. 
We assume that $\limsup_{p\to \infty} M_r<\infty$ for all $r$. 
Similar to Bai and Saranadasa \cite{Bai:1996} and Aoshima and Yata \cite{Aoshima:2013}, we assume that 
\begin{description}
\item[(A-i)]\ $E(w_{rj}^2w_{s j}^2)=E(w_{rj}^2)E(w_{sj}^2)=1$ and $E( w_{rj} w_{s j}w_{tj}w_{u j})=0$\\ for all $r \neq s,t,u$. 
\end{description}
We assume the following assumption instead of (A-i) as necessary:
\begin{description}
\item[(A-ii)]\ $E(w_{r_1j}^{\alpha_1}w_{r_2j}^{\alpha_2}\cdots w_{r_{v} j}^{\alpha_v})=E(w_{r_1j}^{\alpha_1})E(w_{r_2j}^{\alpha_2})\cdots E(w_{r_vj}^{\alpha_v})$ \\
for all $r_1\neq r_2 \neq  \cdots \neq r_v \in [1,q]$ and $\alpha_i\in [1,4]$, $i=1,...,v$, where $v \le 8$ and $\sum_{i=1}^v\alpha_i\le 8$.
\end{description} 
See Chen and Qin \cite{Chen:2010} and Zhong and Chen \cite{Zhong:2011} about (A-ii). 
Note that (A-ii) implies (A-i). 
When $\bx_j$ is Gaussian, it holds that $\bGamma=\bH \bLam^{1/2}$ and $\bw_{j}=\bz_{j}$ in (\ref{2.1}).
Note that (A-ii) is naturally satisfied when $\bx_j$ is Gaussian because the elements of $\bz_{j}$ are independent and $M_r=2$ for all $r$. 
We assume the following assumption for $\bSigma_i$s as necessary: 
\begin{description}
  \item[(A-iii)]\quad  $\displaystyle \min_{i=1,2}\Big\{  \frac{\tr(\bSigma_i^4)}{\tr(\bSigma_i^2)^2 } \Big\} \to 0$ as $p\to \infty$.
\end{description}
We note that if $p_i \to \infty$ and $\tr(\bSigma_i^4)/\tr(\bSigma_i^2)^2\to 0$ as $p\to \infty$, (A-iii) holds even when $p_{i'}$ is fixed for $i'\neq i$.
Also, note that ``$\tr(\bSigma_i^{4})/\tr(\bSigma_i^{2})^{2}\to 0$ as $p\to \infty$" is equivalent to ``$\lambda_{\max}(\bSig_i)/\tr(\bSigma_i^{2})^{1/2}\to 0$ as $p\to \infty$".
Here, $\lambda_{\max}(\bSig_i)$ denotes the largest eigenvalue of $\bSig_i$. 
Let $m=\min\{p,n\}$ and $\Delta=\tr(\bSig_{*}\bSig_{*}^T)\ (=||\bSig_{*} ||_F^2)$, where $||\cdot||_F$ is the Frobenius norm. 
We note that $\Delta=0$ is equivalent to $\bP=\bO$. 
We assume one of the following two assumptions as necessary:
\begin{description}
  \item[(A-iv)]\quad  $\displaystyle \frac{ \tr(\bSigma_1^2)\tr(\bSigma_2^2) }{n^2 \Delta^2 }\to 0$ as $m\to \infty$; 
  \item[(A-v)]\quad  $\displaystyle \limsup_{m\to \infty}\Big\{ \frac{n^2 \Delta^2 }{ \tr(\bSigma_1^2)\tr(\bSigma_2^2)} \Big\}<\infty$.
\end{description}
Note that (A-v) holds under the null hypothesis $H_0$ in (\ref{1.1}). 
%%%%%%%%%%%%%%%%%%%%%%%%%%%
\section{ECDM methodology}
%%%%%%%%%%%%%%%%%%%%%%%%%%%
Yata and Aoshima \cite{Yata:2013} developed the ECDM methodology that is an extension of the CDM methodology given by Yata and Aoshima \cite{Yata:2010}. 
One of the advantages of the ECDM methodology is to produce an unbiased estimator having small asymptotic variance at a low computational cost.
See Section 2.5 of Yata and Aoshima \cite{Yata:2013} for the details. 
In this section, we give a test statistic for (\ref{1.1}) by the ECDM methodology.
%%%%%%%%%%%%%%%%%%%%%%%%%%%%%%%%%%%%%%%% 
\subsection{Unbiased estimator by ECDM}
%%%%%%%%%%%%%%%%%%%%%%%%%%%%%%%%%%%%%%%%
We consider an unbiased estimator of $\Delta$ by the ECDM methodology. 
Let $n_{(1)}=\lceil  n/2 \rceil $ and $n_{(2)}=n-n_{(1)}$, where $ \lceil x \rceil$ denotes the smallest integer $\ge x$. 
Let 
\begin{align*}
&\bV_{n(1) (k)}=
\begin{cases}
\{ \lfloor k/2 \rfloor -n_{(1)}+1,..., \lfloor k/2 \rfloor \} & \mbox{if }\ \lfloor k/2 \rfloor \ge n_{(1)}, \\
\{ 1,...,\lfloor k/2 \rfloor \}\cup \{\lfloor k/2 \rfloor +n_{(2)}+1,...,n \}
  & \mbox{otherwise};
\end{cases} \\
&\bV_{n(2)(k)}=
\begin{cases}
\{ \lfloor  k/2 \rfloor +1,...,\lfloor k/2 \rfloor +n_{(2)}  \} & \mbox{if }\ \lfloor k/2 \rfloor \le n_{(1)},\\
\{1,...,\lfloor k/2 \rfloor -n_{(1)} \} \cup  \{ \lfloor k/2 \rfloor +1,...,n \}   & \mbox{otherwise}
\end{cases}
\end{align*}
for $k=3,...,2n-1$, where $\lfloor x \rfloor $ denotes the largest integer $\le x$. 
Let $\#\bA$ denote the number of elements in a set $\bA$. 
Note that $\#\bV_{n(l)(k)}=n_{(l)}$, $l=1,2$, $\bV_{n(1)(k)} \cap  \bV_{n(2)(k)}=\bem$ and $\bV_{n(1)(k)} \cup \bV_{n(2)(k)}=\{1,...,n\}$ for $k=3,...,2n-1$. 
Also, note that 
\begin{equation}
i\in \bV_{n(1) (i+j)}\quad \mbox{and}\quad j\in \bV_{n(2) (i+j)}\quad \mbox{for $i<j\ (\le n)$}. \label{3.1}
\end{equation}
Let 
$$
\overline{\bx}_{l(1)(k)}=n_{(1)}^{-1} \sum_{j \in \bV_{n(1)(k)}} \bx_{lj} \quad \mbox{and} \quad
\overline{\bx}_{l(2)(k)}=n_{(2)}^{-1} \sum_{j\in \bV_{n(2)(k)}} \bx_{lj},\ \ \mbox{$l=1,2$}
$$
for $k=3,...,2n-1$. 
Let 
$$
\widehat{\Delta}_{ij}=(\bx_{1i}-\overline{\bx}_{1(1)(i+j)})^T(\bx_{1j}-\overline{\bx}_{1(2)(i+j)})(\bx_{2i}-\overline{\bx}_{2(1)(i+j)})^T(\bx_{2j}-\overline{\bx}_{2(2)(i+j)})
$$ 
for all $i< j\ (\le n)$. 
Then, from (\ref{3.1}), we emphasize the following facts:
\begin{align*}
\mbox{(i)}\ &\mbox{$\bx_{1i}-\overline{\bx}_{1(1)(i+j)}$ and $\bx_{1j}-\overline{\bx}_{1(2)(i+j)}$ are independent};\\ 
\mbox{(ii)}\ &\mbox{$\bx_{2i}-\overline{\bx}_{2(1)(i+j)}$ and $\bx_{2j}-\overline{\bx}_{2(2)(i+j)}$ are independent};\\ 
\mbox{(iii)}\ &E(\widehat{\Delta}_{ij})= \Delta
\{(n_{(1)}-1)(n_{(2)}-1)\}/(n_{(1)}n_{(2)})
\end{align*} 
for all $i< j\ (\le n)$.
Let $u_n=n_{(1)}n_{(2)}/\{(n_{(1)}-1)(n_{(2)}-1) \}$. 
We propose an unbiased estimator of $\Delta$ by
\begin{align}
\widehat{T}_{n}&=  \frac{2u_n}{n(n-1) } \sum_{i<j}^{n}\widehat{\Delta}_{ij}. \notag 
\end{align}
%%%%%%%%%%%%%%%%%%
{\bf Remark 1}. \ 
One can save the computational cost of $\widehat{T}_{n}$ by using previously calculated 
$\overline{\bx}_{1(i)(k)}$ and $\overline{\bx}_{2(i)(k)},\ k=3,...,2n-1;\ i=1,2$. 
Then, the computational cost of $\widehat{T}_{n}$ is of the order, $O(n^2p)$. 
\\
%%%%%%%%%%%%%%%%%%

If one considers a naive estimator of $\Delta$ as $\tr(\bS_{*} \bS_{*}^T)$ having
$\bS_{*}=\sum_{j=1}^n(\bx_{1j}-\overline{\bx}_{1n})(\bx_{2j}-\overline{\bx}_{2n})^T/(n-1)$ with $\overline{\bx}_{in}= n^{-1}\sum_{j=1}^n\bx_{ij},\ i=1,2$, it follows that under (A-i)
$$
E\{\tr(\bS_{*} \bS_{*}^T)\}=\Delta+O\Big(\frac{\tr(\bSig_1)\tr(\bSig_2)}{n}\Big).
$$
Note that the bias term of $\tr(\bS_{*} \bS_{*}^T)$ becomes very large as $p$ increases. 
Srivastava and Reid \cite{Srivastava:2012} considered an estimator of $\Delta$ by 
$$
\widehat{\Delta}_{SR}=\frac{(n-1)^2}{(n-2)(n+1)} \Big( \tr(\bS_{*} \bS_{*}^T)-\frac{\tr(\bS_{1})\tr(\bS_{2})}{n-1} \Big)
$$ 
with $\bS_{i}$s the sample covariance matrices when the underlying distribution is Gaussian.
They showed that $E(\widehat{\Delta}_{SR})=\Delta$. 
However, $\widehat{\Delta}_{SR}$ is very biased without the Gaussian assumption. 
Contrary to that, the proposed estimator, $\widehat{T}_n$, is always unbiased and one can claim that $E(\widehat{T}_{n})=\Delta$ without any assumptions. 
\\[5mm]
%%%%%%%%%%%%%%%%%%
{\bf Remark 2}. \ 
We give the following Mathematica algorithm to calculate $\widehat{T}_n$:
\\
{\bf Input}: Sample size $n$ and $n \times p_i$ data matrices $X[i\ ]$, $i=1,2$, such as 
$X[ i\ ]=(\bx_{i1},...,\bx_{in})^T$.\\
{\bf Mathematica code}: {\small
\begin{itemize}
  \item $n1=$Ceiling$[n/2];$ $n2=n-n1;$ $u=2*n1*n2/((n1 - 1)*(n2 - 1)*n*(n-1))$
  \item V$[1,k_-,X_-]$ :=If [Floor$[k/2]\ge n1$, Take$[X, \{$Floor$[k/2]-n1+1$, Floor$[k/2]\}],$\\
   Join[Take$[X, \{1,$ Floor$[k/2]\}],$ Take$[X,$ \{Floor$[k/2]+n2+1, n\}]\ ]\ ]$
  \item V$[2,k_-,X_-]$ :=If [Floor$[k/2]\le n1$, Take$[X, \{$Floor$[k/2]+1$, Floor$[k/2] + n2 \}]$,\\
  Join$[$Take$[X, \{1,$ Floor$[k/2]-n1\},$ Take$[X, \{$Floor$[k/2] + 1, n\}]\ ]\ ]$
  \item Do$[$M$[i,j,k]=$Mean$[V [j,k,X[i]\ ]$, $\{k,3,2*n-1\},$ $\{i,1,2\}$, $\{j,1,2\}]$
  \item $T=u*$Sum[(Part$[X[1], i]-$M$[1,1,i+j]$).(Part$[X[1], j]-$M$[1,2,i+j]$)\\
  $*$(Part$[X[2], i]-$M$[2,1,i+j]$).(Part$[X[2], j]-$M$[2,2,i+j])$, $\{j, 2, n\},\ \{i, 1, j - 1\}$]
\end{itemize}
}
Then, one obtains $T=\widehat{T}_n$.
%%%%%%%%%%%%%%%%%%%%%%%%%%%%%%%%%%%%%%%%%%%%%%%%%%%%%%
\subsection{Asymptotic properties of $\widehat{T}_n$}
%%%%%%%%%%%%%%%%%%%%%%%%%%%%%%%%%%%%%%%%%%%%%%%%%%%%%%
We first consider the consistency property of $\widehat{T}_{n}$ in the sense that $\widehat{T}_{n}/\Delta=1+o_P(1)$ as $m\to \infty$. 
%%%%%%%%%%%. 
\begin{lem}
Assume (A-i). 
It holds that as $m\to \infty$
\begin{align*}
\Var( \widehat{T}_{n})=\Big\{&4\frac{\tr(\bSig_1 \bSig_{*}\bSig_2\bSig_{*}^T)+\tr\{(\bSig_{*}\bSig_{*}^T)^2\}+\sum_{j=1}^q(M_j-2) (\bgamma_{1j}^T\bSig_{*}\bgamma_{2j})^2 }{n}
 \\
&+2\frac{\tr(\bSig_1^2)\tr(\bSig_2^2)+\Delta^2}{n^2}\Big\}\{1+o(1)\}+O\Big(\frac{\{\tr(\bSig_1^4)\tr(\bSig_2^4)\}^{1/2}}{n^2}\Big).
\end{align*}
\end{lem}
%%%%%%%%%%%%%%%
\noindent
{\bf Remark 3.}  
When the underlying distribution is Gaussian and $\bSig_*=\bO$, Srivastava and Reid \cite{Srivastava:2012} showed that as $m\to \infty$
$$
\Var( \widehat{\Delta}_{SR})=\frac{2\tr(\bSig_1^2)\tr(\bSig_2^2)}{n^2}\{1+o(1)\}
$$ 
under certain regularity condition which is stronger than (A-iii).
However, as for $\widehat{T}_{n}$, one can claim that Var($\widehat{T}_{n}$) in Lemma 3.1 is asymptotically equivalent to Var($\widehat{\Delta}_{SR}$) under (A-iii) and $\bSig_*=\bO$.
\\
%%%%%%%%%%%%%%%%

Note that $M_r=2$ for all $r$ when the underlying distribution is Gaussian.
From Lemma 3.1, we have the consistency property of $\widehat{T}_{n}$ as follows:
%%%%%%%%%%%
\begin{thm}
Assume (A-i) and (A-iv). Then, it holds that as $m \to \infty$
\begin{align*}
\frac{\widehat{T}_{n}}{\Delta}=1+o_P(1).
\end{align*}
\end{thm}
%%%%%%%%%%%%
The consistency property holds under (A-iv).
When (A-iv) is not met, we consider the asymptotic normality of $\widehat{T}_{n}$. 
Let $\delta=\{2\tr(\bSigma_1^2)\tr(\bSigma_2^2)\}^{1/2}/n$. 
We give the following result. 
%%%%%%%%%%%%
\begin{lem}
Assume (A-i), (A-iii) and (A-v). 
Then, it holds that as $m \to \infty$
\begin{align*}
\frac{\Var( \widehat{T}_{n})}{\delta^2}=1+o(1).
\end{align*}
\end{lem}
%%%%%%%%%%%%

From Lemma 3.2 we have the asymptotic normality of $\widehat{T}_{n}$ as follows: 
%%%%%%%%%%%
\begin{thm}
Assume (A-ii), (A-iii) and (A-v). 
Then, it holds that as $m \to \infty$
\begin{equation}
\frac{\widehat{T}_{n}-\Delta }{ \sqrt{\Var( \widehat{T}_{n})} }=\frac{\widehat{T}_{n}-\Delta }{\delta}+o_P(1)\Rightarrow N(0,1), 
\notag
\end{equation}
where ``$\Rightarrow$" denotes the convergence in distribution and $N(0,1)$ denotes a random variable distributed as the standard normal distribution. 
\end{thm} 
%%%%%%%%%%%%%%%%%%%%%%%%%%%%%%%%%%%%%%%%%%%%
\subsection{Estimation of $\tr(\bSig_i^2)$}
%%%%%%%%%%%%%%%%%%%%%%%%%%%%%%%%%%%%%%%%%%%%
Since $\tr(\bSig_i^2)$s are unknown in $\delta$, it is necessary to estimate $\tr(\bSig_i^2)$s for constructing a test for (\ref{1.1}).
Yata and Aoshima \cite{Yata:2013} gave an estimator of $\tr(\bSigma_i^2),\ i=1,2,$ by
$$
W_{in}=\frac{2u_n}{n(n-1)}\sum_{r<s}^{n} 
{ \big\{ (\bx_{ir}-\overline{\bx}_{i(1)(r+s)})^T(\bx_{is}-\overline{\bx}_{i(2)(r+s)} )\big\}^2}.
$$ 
Note that $E(W_{in})=\tr(\bSigma_i^2)$. 
From Lemma 3.1, we have the following result.
%%%%%%%%%%%%
\begin{lem}
Assume (A-i). 
Then, it holds as $m\to \infty$ that for $i=1,2$
\begin{align*}
&\Var\Big( \frac{W_{in}}{\tr(\bSig_i^2)}\Big)\\
&=
\Big\{\frac{4}{n\tr(\bSig_i^2)^2}\Big( 2\tr(\bSig_i^4)+\sum_{j=1}^q(M_j-2) (\bgamma_{ij}^T\bSig_{i}\bgamma_{ij})^2  \Big)+
\frac{4}{n^2}\Big\}\{1+o(1)\}\to 0.
\end{align*}
\end{lem}
%%%%%%%%%%%%%
\noindent
{\bf Remark 4}. 
In Section 2.5 of Yata and Aoshima \cite{Yata:2013}, they compared $W_{in}$ with other estimators of $\tr(\bSigma_i^2)$ theoretically and computationally. 
They showed that $W_{in}$ has small asymptotic variance at a low computational cost. 
\\
%%%%%%%%%%%%%

Let $\widehat{\delta}=(2 W_{1n}W_{2n})^{1/2}/n$. 
%Note that $E(\widehat{\delta}^2)=\delta^2$. 
Then, by combining Theorem 3.2 with Lemma 3.3, we have the following result.
%%%%%%%%%%%%
\begin{cor}
Assume (A-ii), (A-iii) and (A-v). 
It holds that as $m \to \infty$
$$
\frac{\widehat{T}_{n}-\Delta }{\widehat{\delta} }\Rightarrow N(0,1).
$$
\end{cor}
%%%%%%%%%%%%%
Now, we considered an easy example such as $p_1=p_2$, $\bmu=\bze$, $\bSigma_1=(0.3^{|i-j|^{1/3}})$, $\bSigma_2=(0.4^{|i-j|^{1/3}})$ and $\bGamma=\bH \bLam^{1/2}$. 
Let $\bSig_i=\bH_i\bLam_i \bH_i^T$ for $i=1,2$, where $\bLam_i=\mbox{diag}(\lambda_{i1},...,\lambda_{ip_i})$ with eigenvalues, $\lambda_{i1}\ge \cdots \ge \lambda_{ip_i}\ (\ge 0)$, and $\bH_i$ is an orthogonal matrix of the corresponding eigenvectors. 
We considered two cases: (a) $\Delta=0$ ($\bx_{1j}=\bH_1\bLam_1^{1/2}( w_{1j},...,w_{p_1j})^T$ and
$\bx_{2j}=\bH_2\bLam_2^{1/2}( w_{p_1+1j},...,w_{pj})^T$), and (b) 
$\Delta=\lambda_{13}\lambda_{23}$ ($\bx_{1j}=\bH_1\bLam_1^{1/2}( w_{1j},...,$
$w_{p_1j})^T$ and
$\bx_{2j}=\bH_2\bLam_2^{1/2}(w_{p_1+1 j},w_{p_1+2j},w_{3j},w_{p+4j},...,w_{pj})^T$). 
Here, $\bx_{j},\ j=1,...,n,$ were generated independently from a pseudorandom normal distribution with mean vector zero and covariance matrix $\bSig$ for each case of $(p,n)=(10,25)$, $(200, 50)$ and $(4000, 150)$. 
Note that (A-ii), (A-iii) and (A-v) hold from the fact that $\Delta=O(1)$.
In Figure 2, we gave two histograms of 2000 independent outcomes of $\widehat{T}_{n}/\widehat{\delta}$ 
for (a) and (b) in each case of $(p,n)$ together with probability densities of $N(0,1)$ and $N(\Delta/\delta,1)$. 
From Corollary 3.1, we expected that $\widehat{T}_{n}/\widehat{\delta}$ is close to $N(0,1)$ when $\Delta=0$ and $N(\Delta/\delta,1)$ when $\Delta \neq 0$. 
When $(p,n)=(10,25)$, the histograms appear far from the probability densities. 
When $(p,n)=(200, 50)$, the histogram for (a) fits well the probability density of $N(0,1)$. 
However, the histogram for (b) is still far from the probability density of $N(\Delta/\delta,1)$. 
This is because the convergence in Lemma 3.2 is slow for $\Delta\neq 0$ compared to $\Delta=0$. 
As expected, both the histograms fit well the probability densities when $(p,n)=(4000, 150)$. 
For other simulation settings such as $p_1=p-1$ and $p_2=1$, see Section 2 of Yata and Aoshima \cite{Yata:2013}. 
\begin{figure}
\begin{center}
%%%%%%%%%%%%%%%%%%%%%%%%%%%%%%%%
\includegraphics[scale=0.35]{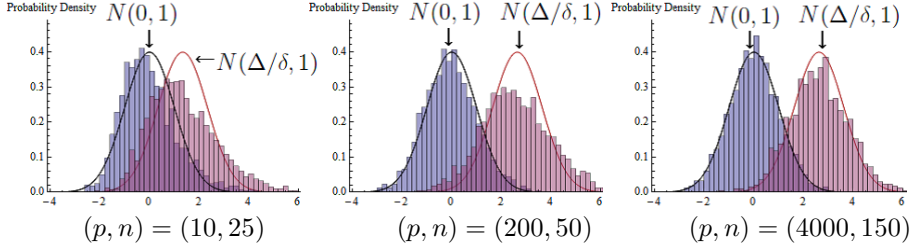} 
\\[-1mm]
{\small 
\quad \quad \quad $(p,n)=(10,25)$\quad \quad \quad \quad \quad $(p,n)=(200, 50)$ \quad \quad \quad \ $(p,n)=(4000, 150)$}
\caption{The solid lines are probability densities of $N(0,1)$ and $N(\Delta/\delta,1)$.  
The histograms of $\widehat{T}_{n}/\widehat{\delta}$ for cases of (a) $\Delta=0$ and (b) $\Delta \neq 0$ fit the solid lines with increasing dimension and sample size: $(p,n)=(10,25)$, $(200,50)$ and $(4000,150)$.} 
\end{center}
\end{figure}
%%%%%%%%%%%%%%%%%%%%%%%%%%%%%%%%%%%%%%%%%%%%%%%%
\section{Test of high-dimensional correlations}
%%%%%%%%%%%%%%%%%%%%%%%%%%%%%%%%%%%%%%%%%%%%%%%%
In this section, we propose a test procedure for (\ref{1.1}) in high-dimensional settings. 
%%%%%%%%%%%%%%%%%%%%%%%%%%%%%%%%%%%%%%%%%%%%
\subsection{Test procedure for (\ref{1.1})}
%%%%%%%%%%%%%%%%%%%%%%%%%%%%%%%%%%%%%%%%%%%%
Let $\alpha \in (0,1/2)$ be a prespecified constant. 
From Corollary 3.1, we test (\ref{1.1}) by 
\begin{align}
\mbox{rejecting}\ H_0 \Longleftrightarrow  \frac{\widehat{T}_{n} }{\widehat{\delta}}
> z_{\alpha}, 
\label{4.1}
\end{align}
where $z_{\alpha}$ is a constant such that $P\{N(0,1)>z_{\alpha}\}=\alpha$. 
Then, we have the following result.
%%%%%%%%%%%%
\begin{thm}
Under (A-ii) and (A-iii), the test by (\ref{4.1}) has that as $m\to \infty$
\begin{equation}
\mbox{size}=\alpha+o(1) \quad \mbox{and}\quad \mbox{power}(\Delta_\star)-\Phi \Big(\frac{\Delta_\star}{\delta}-z_{\alpha} \Big)=o(1),
\notag
\end{equation}
where $\Phi(\cdot)$ denotes the c.d.f. of $N(0,1)$ and power($\Delta_\star$) denotes the power when $\Delta=\Delta_\star$ for given $\Delta_\star(>0)$. 
\end{thm}
%%%%%%%%%%%%%%%
When (A-iv) is met, we have the following result from Theorem 3.1. 
%%%%%%%%%%%%
\begin{cor}
Assume (A-i). 
Assume (A-iv) under $H_1$. 
Then, the test by (\ref{4.1}) has for any $\Delta(>0)$ that as $m\to \infty$
$$
\mbox{Power}(\Delta)=1+o(1).
$$
\end{cor}
%%%%%%%%%%%%
\noindent
{\bf Remark 5}. 
Let
\begin{align*}
K=\Big\{&4\frac{\tr(\bSig_1 \bSig_{*}\bSig_2\bSig_{*}^T)+\tr\{(\bSig_{*}\bSig_{*}^T)^2\}+\sum_{j=1}^q(M_j-2) (\bgamma_{1j}^T\bSig_{*}\bgamma_{2j})^2}{n}\\
&+2\frac{\tr(\bSig_1^2)\tr(\bSig_2^2)+\Delta^2}{n^2}\Big\}^{1/2}.
\end{align*}
Then, from Lemma 3.1, it holds that $\Var(\widehat{T}_n)/K^2 \to 1$ as $m\to \infty$ under (A-i) and (A-iii). 
Hence, from Theorem 3.2, one may write the power in Theorem 4.1 as 
$$
power(\Delta_\star)-\Phi \Big(\frac{\Delta_\star}{K}-\frac{z_{\alpha} \delta}{K} \Big)=o(1). 
$$
%%%%%%%%%%%%%%%%%%%%%%%%
\subsection{Simulation}
%%%%%%%%%%%%%%%%%%%%%%%%
In order to study the performance of the test by (\ref{4.1}), we used computer simulations. 
We set $\alpha=0.05$, $p_1=p_2$, $\bmu=\bze$, $\bSig_1=\bB(0.3^{|i-j|^{1/3}})\bB$, $\bSig_2=\bB(0.4^{|i-j|^{1/3}})\bB$ and $\bGamma=\bH \bLam^{1/2}$, where
$$
\bB=\mbox{diag}[ \{0.5+1/(p_1+1)\}^{1/2},...,\{0.5+p_1/(p_1+1)\}^{1/2} ]. 
$$
Note that $\tr(\bSig_i)=p_i\ (i=1,2)$. 
We set (a) $\Delta=0$ and (b) $\Delta=\lambda_{13}\lambda_{23}$ that are the same settings as in Figure 2. 
We considered three distributions for $\bx_{j}$s: 
(I) $ N_p(\bze, \bSig)$,
(II) $w_{rj}=(v_{rj}-1)/{2}^{1/2}$ $(r=1,...,q)$ in which $v_{rj}$s are i.i.d. as the chi-squared distribution with $1$ degree of freedom 
and  
(III) $\bw_j$s are i.i.d. as $p$-variate $t$-distribution, $t_p(\nu)$, with mean zero, covariance matrix $\bI_p$ and degrees of freedom $\nu=10$. 
Note that (A-ii) is met in (I) and (II). 
However, (A-i) (or (A-ii)) is not met in (III). 
We set $p=2^s\  (s=4,...,11)$ and $n=4 \lceil p_1^{1/2} \rceil$.
We note that (A-iii) and (A-v) hold for (a) and (b).
We compared the performance of $\widehat{T}_n$ with $\widehat{\Delta}_{SR}/\widehat{\delta}_{SR}$ by Srivastava and Reid \cite{Srivastava:2012}, where 
$\widehat{\delta}_{SR}=\{2W_{1(SR)}W_{2(SR)}\}^{1/2}/n$ and $W_{i(SR)}=[(n-1)^2/\{(n-2)(n+1) \}]\{\tr(\bS_i^2)-\tr(\bS_i)^2/(n-1)\}$, $i=1,2$. 
They showed that $\widehat{\Delta}_{SR}/\widehat{\delta}_{SR}$ has the asymptotic normality as $m\to\infty$ when the underlying distribution is Gaussian and $\Delta=0$. 
Also, note that $E(\widehat{\Delta}_{SR})=\Delta$ only under the Gaussian assumption. 
Contrary to that, from Corollary 3.1, $\widehat{T}_n/\widehat{\delta}$ has the asymptotic normality as $m\to\infty$ even for non-Gaussian situations and $\Delta\neq 0$. 
Also, one can claim that $E(\widehat{T}_{n})=\Delta$ without any assumptions such as (A-i).

In Figure 3, we summarized the findings obtained by averaging the outcomes from 4000 $(=R,$ say) replications for (I) to (III).
Here, the first $2000$ replications were generated for (a) when $\Delta=0$ and the last $2000$ replications were generated for (b) when $\Delta\neq 0$. 
We defined $P_{r}=1\ (\mbox{or}\ 0)$ when $H_0$ was falsely rejected (or not) for $r=1,...,2000$, and $H_1$ was falsely rejected (or not) for $r=2001,...,4000$.  
We gave $\overline{\alpha}=(R/2)^{-1} \sum_{r=1}^{R/2}P_{r}$ to estimate the size in the left panels and $1-\overline{\beta}=1-(R/2)^{-1} \sum_{r=R/2+1}^{R}P_{r}$ to estimate the power in the right panels. 
Their standard deviations are less than $0.011$. 
Let $L=\Phi ( \Delta/K-z_{\alpha} \delta/K)$. 
From Theorem 4.1 in view of Remark 5, we expected that $\overline{\alpha}$ and $1-\overline{\beta}$ for (\ref{4.1}) are close to $0.05$ and $L$, respectively. 
In Figure 4, we gave the averages (in the left panels) and the sample variances (in the right panels) of $\widehat{T}_n/\Delta$ and $ \widehat{\Delta}_{SR}/\Delta$ by the outcomes for (b) when $\Delta\neq 0$ in cases of (I) to (III). 
From Remark 5, the asymptotic variance for $\widehat{T}_n/\Delta$ was given by $K^2/\Delta^2$. 

From Figures 3 and 4, we observed that $ \widehat{\Delta}_{SR}$ gives good performances for the Gaussian case. 
However, for non-Gaussian cases such as (II) and (III), $\widehat{\Delta}_{SR}$ seems not to give a preferable performance. 
Especially, it gave quite bad performances for (III). 
That is probably because (A-i) (or (A-ii)) is not met in (III). 
On the other hand, $\widehat{T}_n$ gave adequate performances for high-dimensional cases even in the non-Gaussian situations. 
We observed that $\widehat{T}_n$ is quite robust against other non-Gaussian situations as well. 
Hence, we recommend to use $\widehat{T}_n$ for the test of (\ref{1.1}) and for the estimation of $\Delta$. 
%%%%%%%%%%%%%%
\begin{figure}
\begin{center}
\includegraphics[scale=0.48]{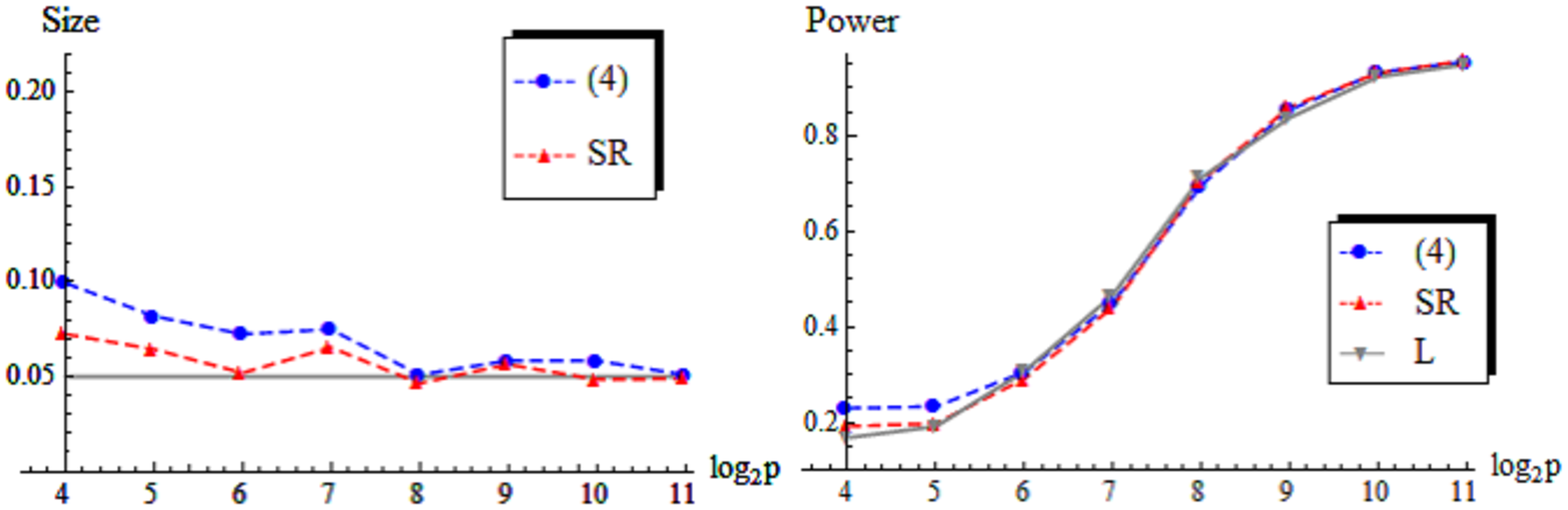} \\[-1mm]
(I) $ N_p(\bze, \bSig)$. \\[3mm]
\includegraphics[scale=0.48]{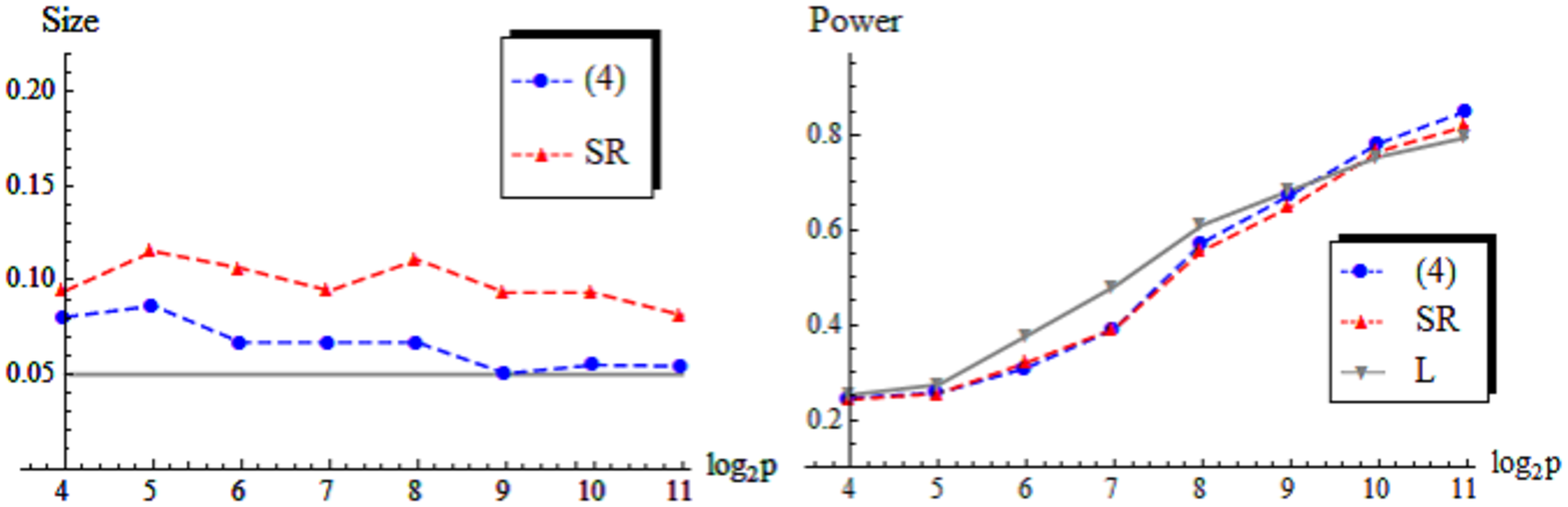} \\[-1mm]
(II) The chi-squared distribution with $1$ degree of freedom. \\[3mm] 
\includegraphics[scale=0.48]{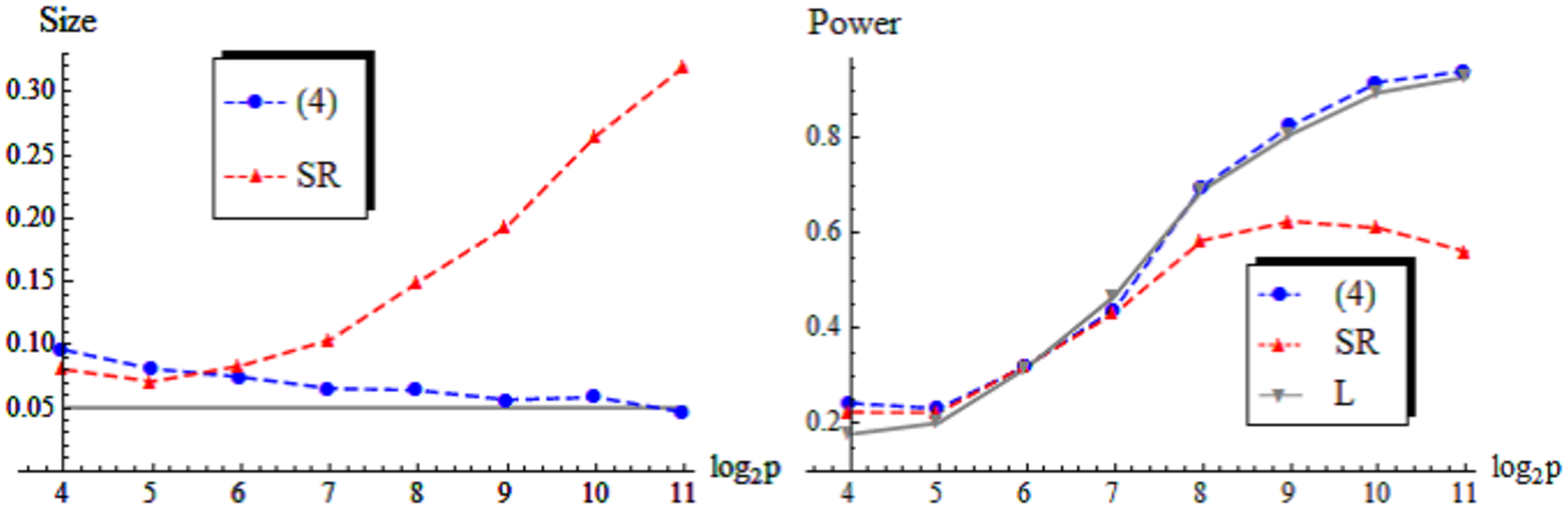} \\[-1mm]
(III) $t_p(10)$. 
\end{center}
\caption{The values of $\overline{\alpha}$ are denoted by the dashed lines in the left panels and the values of $1-\overline{\beta}$ are denoted by the dashed lines in the right panels for the tests by (\ref{4.1}) and $\widehat{\Delta}_{SR}/\widehat{\delta}_{SR}$ (SR) in cases of (I) to (III). 
The asymptotic powers were given by $L=\Phi ( \Delta/K-z_{\alpha} \delta/K)$ which was denoted by the solid lines in the right panels.  
}
\end{figure}
%%%%%%%%%%%%%%%
\begin{figure}
\begin{center}
\includegraphics[scale=0.48]{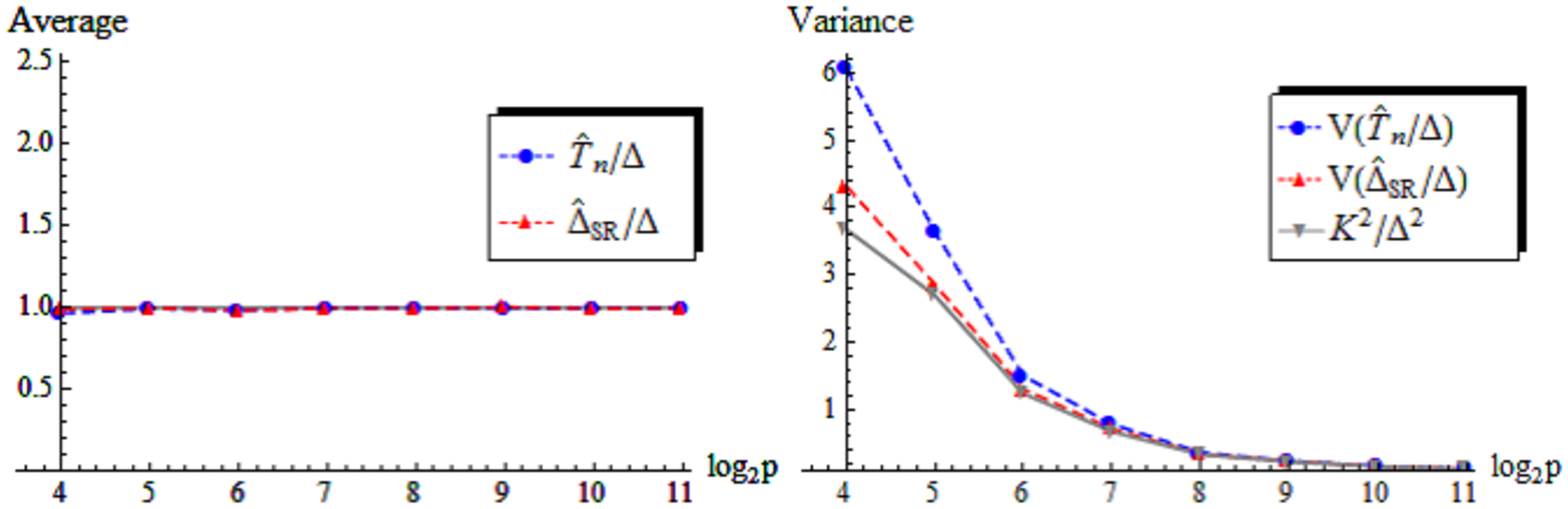} \\[-1mm]
(I) $ N_p(\bze, \bSig)$. \\[3mm]
\includegraphics[scale=0.48]{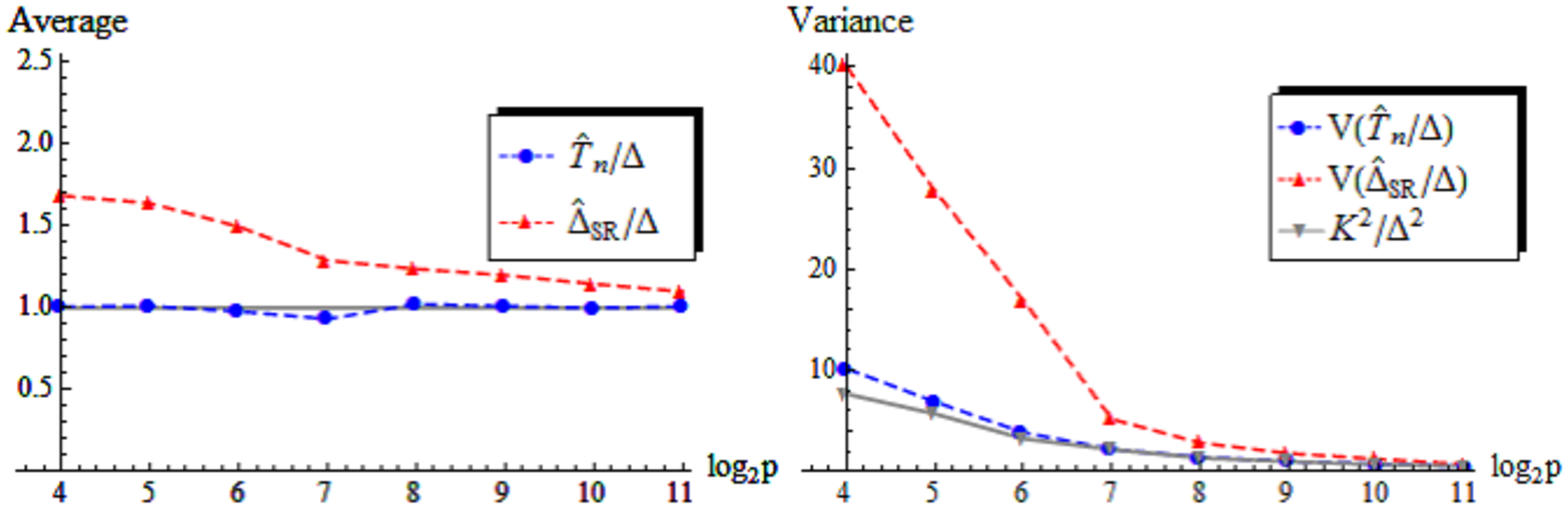} \\[-1mm]
(II) The chi-squared distribution with $1$ degree of freedom. \\[3mm] 
\includegraphics[scale=0.48]{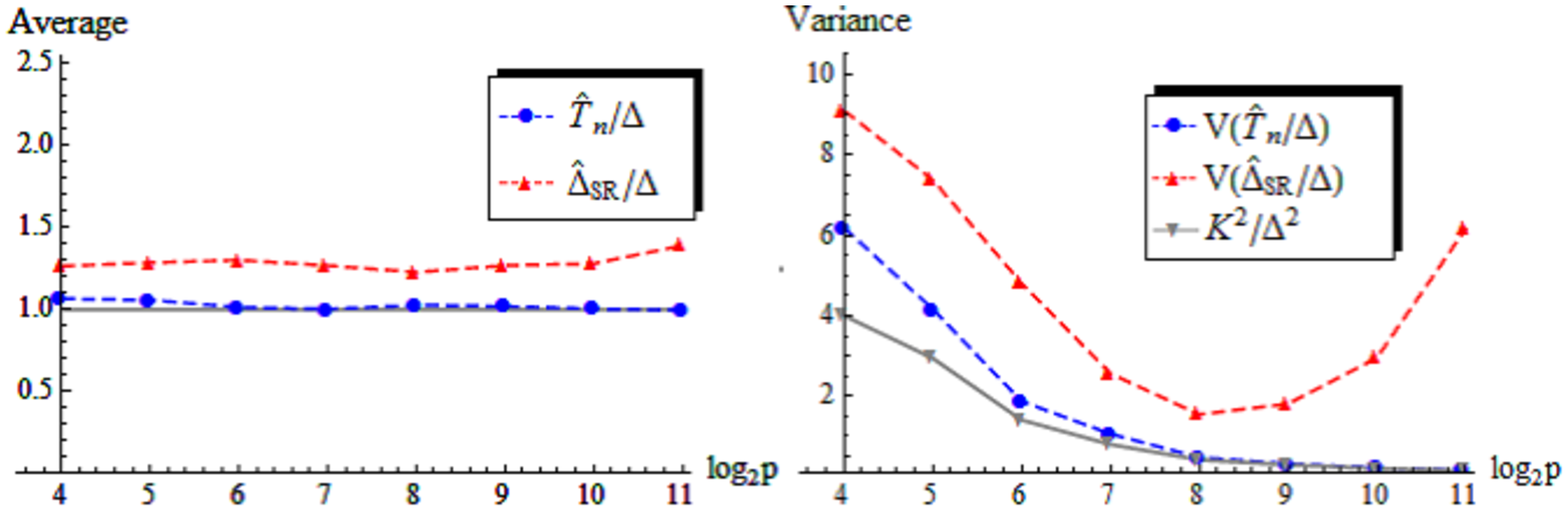} \\[-1mm]
(III) $t_p(10)$.
\end{center}
\caption{The averages of $\widehat{T}_n/\Delta$ and $ \widehat{\Delta}_{SR}/\Delta$ are denoted by the dashed lines in the left panels and their sample variances, 
$V(\widehat{T}_n/\Delta)$ and $V( \widehat{\Delta}_{SR}/\Delta)$, are denoted by the dashed lines in the right panels in cases of (I) to (III). 
The asymptotic variance of $\widehat{T}_n/\Delta$ was given by $K^2/\Delta^2$ which was denoted by the solid lines in the right panels.}
\end{figure}
%%%%%%%%%%%%%%%%%%%%%%%
\section{Applications}
%%%%%%%%%%%%%%%%%%%%%%%
In this section, we give several applications of the results in Section 3.
\subsection{Confidence interval for $\Delta$}
We construct a confidence interval for $\Delta$ by 
\begin{align*}
&I=[\max\{\widehat{T}_{n}-z_{\alpha/2}\widehat{\delta},\ 0 \},\ \widehat{T}_{n}+z_{\alpha/2}\widehat{\delta}],
\end{align*}
where $\alpha \in(0,1)$. 
Then, from Corollary 3.1, it holds that as $m\to \infty$
$$
P( \Delta  \in I)= 1-\alpha+o(1)%\quad \mbox{and} \quad P(\Delta  \in I_L)= 1-\alpha+o(1)
$$
under (A-ii), (A-iii) and (A-v). 
Hence, one can estimate $\Delta$ by $I$. 
If one considers $\bSig_{0}$ as a candidate of $\bSig_*$, one can check whether $\bSig_{0}$ is a valid candidate or not according as $||\bSig_{0} ||_F^2 \in I$ or not.
%%%%%%%%%%%%%%%%%%%%%%%%%%%%%%%%%%%%%%%%%%%%%%%%%%
\subsection{Checking whether (A-iv) holds or not}
%%%%%%%%%%%%%%%%%%%%%%%%%%%%%%%%%%%%%%%%%%%%%%%%%%
As discussed in Section 3, $\widehat{T}_{n}$ holds the consistency property when (A-iv) is met, and $\widehat{T}_{n}$ holds the asymptotic normality when (A-v) is met.
Here, we propose a method to check whether (A-iv) holds or not.

Let $\widehat{\kappa}=W_{1n}W_{2n}/(n\widehat{T}_n)^2$. 
We have the following result. 
%%%%%%%%%%%%
\begin{pro}
Assume (A-i). 
It holds that as $m\to \infty$. 
$$
\widehat{\kappa}=o_P(1) \quad \mbox{under (A-iv)};
\quad 
\widehat{\kappa}^{-1}=O_P(1)\quad \mbox{under (A-v)}.
$$
\end{pro}
%%%%%%%%%%%%
From Proposition 5.1, one can distinguish (A-iv) and (A-v). 
If $\widehat{\kappa}$ is sufficiently small, one may claim (A-iv), otherwise (A-v). 
%%%%%%%%%%%%%%%%%%%%%%%%%%%%%%%%%%%%%%%%%%%%%%
\subsection{Estimation of the RV-coefficient}
Let $\rho=\Delta/\{{\tr(\bSig_1^2)\tr(\bSig_2^2)}\}^{1/2}$.
Here, $\rho$ is the (population) RV-coefficient which is a multivariate generalization of the squared Pearson correlation coefficient. 
Note that $\rho \in [0,1]$. 
See Robert and Escoufier \cite{Robert:1974} for the details. 
Smilde et al. \cite{Smilde:2009} considered the RV-coefficient for high-dimensional data. 

Let $\widehat{\rho}=\widehat{T}_n/(W_{1n}W_{2n})^{1/2}$. 
Then, we have the following result. 
%%%%%%%%%%%%
\begin{pro}
Assume (A-i). 
It holds that as $m\to \infty$
$$
\widehat{\rho}=\rho+O_P(1/n+\rho/n^{1/2})+O_P\Big\{ \Big(\frac{\tr(\bSig_1 \bSig_{*}\bSig_2\bSig_{*}^T)}{\tr(\bSig_1^2)\tr(\bSig_2^2)n}\Big)^{1/2} \Big\}=O_P(n^{-1/2}).
$$
\end{pro}
%%%%%%%%%%%%
Thus, one can estimate the RV coefficient by $\widehat{\rho}$ for high-dimensional data. 
%%%%%%%%%%%%%%%%%%%%%%%%%%%%%%%%%%%%%%%%%%%%%%%%%%%%%%%%%%%%
\subsection{Test of high-dimensional covariance structures}
%%%%%%%%%%%%%%%%%%%%%%%%%%%%%%%%%%%%%%%%%%%%%%%%%%%%%%%%%%%%
We consider testing 
\begin{align}
H_0: \bSig_{*}=\bSig_{0} \quad \mbox{vs.}\quad H_1:\bSig_{*}\neq \bSig_{0}, \label{5.1}
\end{align}
where $\bSig_{0}$ is a candidate covariance structure. 
Let $\Delta_0=||\bSig_*-\bSig_0||_F^2$ and 
\begin{align*}
\widehat{\Delta}_{ij,0}=&u_n\widehat{\Delta}_{ij}
-n_{(1)}(\bx_{1i}-\overline{\bx}_{1(1)(i+j)})^T\bSig_{0}(\bx_{2i}-\overline{\bx}_{2(1)(i+j)})/(n_{(1)}-1)\\
&-n_{(2)}(\bx_{1j}-\overline{\bx}_{1(2)(i+j)})^T\bSig_{0}(\bx_{2j}-\overline{\bx}_{2(2)(i+j)})/(n_{(2)}-1),
\end{align*}
where $u_n=n_{(1)}n_{(2)}/\{(n_{(1)}-1)(n_{(2)}-1) \}$.
Note that $E(\widehat{\Delta}_{ij,0})=||\bSig_*||_F^2-2\tr(\bSig_*^T\bSig_0)=\Delta_0-||\bSig_{0}||_F^2$. 
Then, we consider a test statistic for (\ref{5.1}) by
\begin{align}
\widehat{T}_{n,0}&=\frac{2}{n(n-1) } \sum_{i<j}^{n}\widehat{\Delta}_{ij,0}+||\bSig_{0}||_F^2.
\notag 
\end{align}
Note that $E(\widehat{T}_{n,0})=\Delta_0$. 
Let $\bSig_{*0}=\bSig_*-\bSig_0$. 
Then, we have the following result.
%%%%%%%%%%%% 
\begin{lem}
\label{lem5.1}
Assume (A-i). Then, it holds that as $m\to \infty$
\begin{align*}
\Var( \widehat{T}_{n})=\Big\{&4\frac{\tr(\bSig_1 \bSig_{*0}\bSig_2\bSig_{*0}^T)+\tr\{(\bSig_{*} \bSig_{*0}^T )^2 \}+\sum_{j=1}^q(M_j-2) (\bgamma_{1j}^T\bSig_{*0}\bgamma_{2j})^2 }{n}
 \\
&+2\frac{\tr(\bSig_1^2)\tr(\bSig_2^2)+\Delta^2}{n^2}\Big\}\{1+o(1)\}+O\Big(\frac{\{\tr(\bSig_1^4)\tr(\bSig_2^4)\}^{1/2}}{n^2}\Big).
\end{align*}
\end{lem}
%%%%%%%%%%%%
From Lemma \ref{lem5.1}, Theorems 3.1 and 3.2, we have the following results.
%%%%%%%%%%%% 
\begin{cor}
Assume (A-i). 
Assume also (A-iv) with $\Delta=\Delta_0$. 
Then, it holds that as $m \to \infty$
\begin{align*}
\frac{\widehat{T}_{n,0}}{\Delta_0}=1+o_P(1).
\end{align*}
\end{cor}
%%%%%%%%%%%%
\begin{cor}
Assume (A-ii), (A-iii) and (A-v). 
Assume also (A-v) with $\Delta=\Delta_0$. 
Then, 
it holds that as $m \to \infty$
$$
\frac{\widehat{T}_{n,0}-\Delta_0 }{\widehat{\delta} }\Rightarrow N(0,1).
$$
\end{cor}
Hence, one can apply $\widehat{T}_{n,0}$ to a test for (\ref{5.1}). 
%%%%%%%%%%%%%%%%%%
\section{Example}
%%%%%%%%%%%%%%%%%%
In this section, we demonstrate how the proposed test procedures perform in actual data analyses by using a microarray data set.
We analyzed gene expression data of Arabidopsis thaliana given by Wille et al. \cite{Wille:2004} in which the data set consists of $118$ samples having $834\ (=p)$ genes: $39\ (=p_1)$ isoprenoid genes and $795\ (=p_2)$ additional genes. 
All the data were logarithmic transformed. 
Wille et al. \cite{Wille:2004} considered a genetic network between the two gene sets. 
By using a graphical Gaussian modeling, they constructed an isoprenoid gene network given in  Figure 2 of \cite{Wille:2004}. 
In Figure 5, we gave the illustration of the isoprenoid gene network and the additional genes. 
We first considered testing (\ref{1.1}) by using (\ref{4.1}). 
See Figure 1 for the illustration. 
Let $\alpha=0.05$. 
We calculated $\widehat{T}_n=352.5$ and $\widehat{\delta}=7.296$, so that $\widehat{T}_n/\widehat{\delta}=48.3$. 
From (\ref{4.1}) and $z_{\alpha}=1.645$, we rejected $H_0$. 
Thus we concluded that two networks have some connections. 
In addition, we calculated $\widehat{\kappa}=0.000214$. 
Thus, with the help of Proposition 5.1 one may conclude that (A-iv) is met, so that the power of the test is $1$ asymptotically and $\widehat{T}_n/\Delta=1+o_P(1)$ from Theorem 3.1 and Corollary 4.1.
Also, with the help of Proposition 5.2 we obtained $\widehat{\rho}=0.579$ as an estimate of the RV-coefficient. 
\begin{figure}
%%%%%%%%%%%%%%%%%%%%%%%%%%%%%%%%
\includegraphics[scale=0.53]{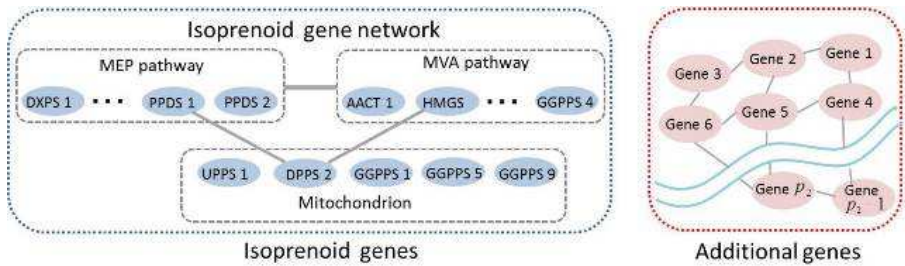}
\caption{
Illustration of the isoprenoid gene network given by Figure 2 in Wille et al. \cite{Wille:2004} and the additional genes, where DXPS$1$, PPDS$1$ and so on are names of genes. 
DPPS$2$ is connected with both MEP pathway and MVA pathway. 
Other genes of mitochondrion are not connected with either MEP pathway or MVA pathway.
}
\end{figure}
%%%%%%%%%%%%%%%%%%%%%%%%%%%%%%%%

Next, we considered testing (\ref{1.1}) between some part of the isoprenoid genes and the additional genes. 
The isoprenoid genes consisted of three types as MEP pathway (19 genes), MVA pathway (15 genes) and mitochondrion (5 genes). 
See \cite{Wille:2004} for the details. 
From Figure 5 we expected that (i) the correlation between DPPS$2$ and the additional genes is high, and 
(ii) the correlation between the genes of mitochondrion (except DPPS$2$) and the additional genes is low. 
We set $\bx_{2j}$ as the additional genes ($p_2=795$). 
We considered three tests for $\bx_{1j}$: (a) the genes of mitochondrion ($p_2=5$); (b) DPPS$2$ ($p_2=1$); and (c) UPPS$1$, GGPPS$1, 5, 9$ ($p_2=4$). 
By using the first 50 samples ($n=50$) of the $118$ samples, we constructed (\ref{4.1}). 
Then, with $\alpha=0.05$, we rejected $H_0$ for (a) since $\widehat{T}_n/\widehat{\delta}=12.27$ and for (b) since $\widehat{T}_n/\widehat{\delta}=13.23$.
On the other hand, we accepted $H_0$ for (c) since $\widehat{T}_n/\widehat{\delta}=1.417$. 

Similar to Section 5 in Yata and Aoshima \cite{Yata:2013}, we considered a high-dimensional linear regression model:
$$
\bY=\bX \bThe +\bE,
$$
where $\bY$ is an $n\times p_2$ response matrix, $\bX$ is an $n\times k$ fixed design matrix, and $\bThe$ is a $k \times p_2$ parameter matrix. 
The $n$ rows of $\bE$ are independent and identically distributed as a $p_2$-variate distribution with mean vector zero. 
Let $\bx_{1j}$ be the $j$th sample of the $35$ isoprenoid genes (except UPPS$1$, GGPPS$1, 5, 9$). 
Let $\bx_{1(j)}=(1, \bx_{1j}^T)^T$, $j=1,...,118$. 
We set $\bY =[\bx_{21},...,\bx_{2n} ]^T$ and $\bX =[\bx_{1(1)},...,\bx_{1(n)}]^T$ with $k=36$. 
We noted that the standard elements of $ \bThe$ are path coefficients from the isoprenoid genes to the additional genes. 
By using the observed samples of size $n=50$ as a training data set, we obtained the least squared estimator of $\bThe$ by 
$\widehat{\bThe}=(\bX^T\bX)^{-1}\bX^T\bY$. 
We investigated prediction accuracy of the regression with $\widehat{\bThe}$ by using the remaining samples of size $68\, (=118-50)$ as a test data set. 
We considered the prediction mean squared error (PMSE) by $E(||{\bx}_{2j}-\widehat{\bThe}^T\bx_{1(j)}||^2 | \widehat{\bThe})$. 
By using the test samples $\bx_{1(j)}$ and ${\bx}_{2j}$, $j=51,...,118$, 
we applied the bias-corrected and accelerated (BCa) bootstrap by Efron \cite{Efron:1987}. 
Then, we constructed $95$\% confidence interval (CI) of the PMSE by $[837.6, 1189.5]$ from 10000 replications.
On the other hand, we considered the PMSE for the full isoprenoid ($39$ genes). 
Then, similar to above, we constructed $95$\% CI of the PMSE by $[1088.7,1581.3]$. 
The PMSE by the $35$ isoprenoid genes is probably smaller than that of the full
isoprenoid genes. 
Thus we conclude that the test procedure by (\ref{4.1}) effectively works for this data set. 
%%%%%%%%%%%%%%%%%
\appendix
\section{Proofs}
\label{app}
%%%%%%%%%%%%%%%%%
Throughout, we assume that $\bmu_1=\bze$ and $\bmu_2=\bze$ without loss of generality. 
Let $\Upsilon=\tr(\bSig_1\bSig_* \bSig_2 \bSig_*^T)$, $\Psi=\tr(\bSig_1^2)\tr(\bSig_2^2)$ and 
$\Omega=\tr(\bSig_1^4)\tr(\bSig_2^4)$. 
Note that
\begin{align}
&\sum_{i=1}^q(\bgamma_{1i}^T\bSig_*\bgamma_{2i} )^2\le \sum_{i,j}^q(\bgamma_{1i}^T\bSig_*\bgamma_{2j})^2=\Upsilon; \notag \\
&\tr\{(\bSig_* \bSig_*^T)^2\}=\sum_{i,j}^q(\bgamma_{1i}^T\bSig_*\bgamma_{2j})(\bgamma_{1j}^T\bSig_*\bgamma_{2i})\le \sum_{i,j}^q(\bgamma_{1i}^T\bSig_*\bgamma_{2j})^2=\Upsilon; \notag\\
&\Delta=\sum_{i,j}^q(\bgamma_{1i}^T\bgamma_{1j}\bgamma_{2i}^T\bgamma_{2j})\le
\Big\{\sum_{i,j}^q(\bgamma_{1i}^T\bgamma_{1j})^2\Big\}^{1/2}\Big\{ \sum_{i,j}^q(\bgamma_{2i}^T\bgamma_{2j})^2\Big\}^{1/2}=\Psi^{1/2}; \notag
\\
&\Upsilon=\sum_{i,j}^q(\bgamma_{1i}^T\bSig_1\bgamma_{1j} )(\bgamma_{2i}^T\bSig_2 \bgamma_{2j})\le 
\Big\{ \sum_{i,j}^q(\bgamma_{1i}^T\bSig_1\bgamma_{1j} )^2\Big\}^{1/2}
\Big\{ \sum_{i,j}^q(\bgamma_{2i}^T\bSig_2\bgamma_{2j} )^2\Big\}^{1/2}\notag\\
&\hspace{4.7cm}=\tr(\bSig_1^4)^{1/2} \tr(\bSig_2^4)^{1/2}=\Omega^{1/2}\le \Psi; \ \mbox{ and} \notag
\\
&\sum_{i=1}^q(\bgamma_{1i}^T\bSig_1\bgamma_{1i})(\bgamma_{2i}^T\bSig_2\bgamma_{2i})
\le \Big\{ \sum_{i=1}^q(\bgamma_{1i}^T\bSig_1\bgamma_{1i} )^2 \Big\}^{1/2}
\Big\{ \sum_{i=1}^q(\bgamma_{2i}^T\bSig_2\bgamma_{2i} )^2 \Big\}^{1/2}\notag \\
&\hspace{3.95cm}\le \Omega^{1/2}\le \Psi 
\label{A.1}
\end{align}
from the fact that $\tr(\bSig_i^4)\le \tr(\bSig_i^2)^2$ for $i=1,2$. 
Then, we note that $K^2=O(\Psi/n^2+\Upsilon/n)$, where $K$ is given in Remark 5.
Let $y_{ij}=u_n\widehat{\Delta}_{ij}-\Delta$ and $\varepsilon_{ij}=\bx_{1i}^T\bx_{1j}\bx_{2i}^T\bx_{2j}-\Delta$ for all $i < j$. 
Note that $\widehat{T}_n-\Delta=2\sum_{i<j}^ny_{ij}/\{n(n-1)\}$.  
Let $\eta_{ij}=\sum_{r\neq s}^q\sum_{t=1}^q \bgamma_{1r}^T\bgamma_{1t}\bgamma_{2s}^T\bgamma_{2t}w_{ri}w_{si}(w_{tj}^2-1)$, 
$\psi_{ij}=\sum_{r,t}^q \bgamma_{1r}^T\bgamma_{1t}\bgamma_{2r}^T\bgamma_{2t}(w_{ri}^2-1)(w_{tj}^2-1)$ and 
$\omega_{ij}=\sum_{r\neq s}^q\sum_{t\neq u}^q \bgamma_{1r}^T\bgamma_{1t}\bgamma_{2s}^T\bgamma_{2u}w_{ri}w_{si}w_{tj}w_{uj}$ for all $i\neq j$. 
Note that $E(\omega_{ij})=0$ for all $i\neq j$ and $E(\omega_{ij}\omega_{i' j})=0$ for all $i\neq i'\neq j$.
Let $U_n=2\sum_{i<j}^n\varepsilon_{ij}/\{n(n-1)\}$, $V_n=2\sum_{i<j}^n \omega_{ij}/\{n(n-1)\}$ and $B=E(\omega_{ij}^2)\ (i \neq j)$.
%Note that $B=\Psi+O(\Upsilon)$ under (A-i). 
Let $\widehat{\bSig}_{*,ij(1)}=n_{(1)}(\bx_{1i}-\overline{\bx}_{1(1)(i+j)})(\bx_{2i}-\overline{\bx}_{2(1)(i+j)})^T/(n_{(1)}-1)$ and 
$\widehat{\bSig}_{*,ij(2)}=n_{(2)}(\bx_{1j}-\overline{\bx}_{1(2)(i+j)})(\bx_{2j}-\overline{\bx}_{2(2)(i+j)})^T/(n_{(2)}-1)$
for all $i<j$. 
%%%%%%%%%%%%%%%%%%%%%%%%%%%%%%%%%%
\begin{proof}[Proof of Lemma 3.1]
%%%%%%%%%%%%%%%%%%%%%%%%%%%%%%%%%%
We write that 
\begin{align}
y_{ij}=&\tr\{(\widehat{\bSig}_{*,ij(1)}-\bSig_*)(\widehat{\bSig}_{*,ij(2)}-\bSig_*)^T \} 
\notag \\
&+\tr(\widehat{\bSig}_{*,ij(1)}\bSig_*^T)+\tr(\widehat{\bSig}_{*,ij(2)}\bSig_*^T)-2\Delta^2; \notag \\
\varepsilon_{ij}=&\omega_{ij}+\eta_{ij}+\eta_{ji}+\psi_{ij}+
\tr(\bx_{1i} \bx_{2i}^T\bSig_*^T)
+\tr(\bx_{1j}\bx_{2j}^T\bSig_*^T)-2\Delta^2
\label{A.2}
\end{align}
for all $i<j$. 
Note that all the terms of $\varepsilon_{ij}$ in (\ref{A.2}) are uncorrelated under (A-i). 
From (\ref{A.1}), it holds that under (A-i) 
$$
E(\psi_{ij}^2)=O\Big(\sum_{r,t}^q (\bgamma_{1r}^T\bgamma_{1t}\bgamma_{2r}^T\bgamma_{2t})^2 \Big)=
O\Big(\sum_{r=1}^q \bgamma_{1r}^T\bSig_1\bgamma_{1r}\bgamma_{2r}^T\bSig_2\bgamma_{2r} \Big)=O(\Omega^{1/2})
$$ 
for all $i\neq j$. 
Similarly, under (A-i), it holds that $E(\eta_{ij}^2)=O(\Omega^{1/2})$ for all $i\neq j$. 
Then, we have that under (A-i)
\begin{align*}
&E(\varepsilon_{ij}^2)=\Psi+\Delta^2+O(\Upsilon+\Omega^{1/2})\quad \mbox{for all $i< j$}; \\
&E(\varepsilon_{ij}\varepsilon_{ik})=E(\varepsilon_{ik}\varepsilon_{jk})=\Var\{\tr(\bx_{1i} \bx_{2i}^T\bSig_*^T)\}
\\
&\hspace{1.4cm}
=\Upsilon+\tr\{(\bSig_* \bSig_*^T)^2\}+\sum_{r=1}^q(M_r-2)(\bgamma_{1r}^T\bSig_*\bgamma_{2r} )^2  \quad \mbox{for all $i< j< k$}; \\
&\mbox{and }\ E(\varepsilon_{ij}\varepsilon_{kl})=0\quad \mbox{for all $i< j$ and $ k< l;\ i \neq j\neq k \neq l$}. 
\end{align*} 
Then, we have that as $m\to \infty$ 
\begin{align}
&\Var( U_{n})=E(U_n^2)=K^2\{1+o(1)\}+O(\Omega^{1/2}/n^2)=O(K^2)
\label{A.3}
\end{align}
under (A-i).
On the other hand, we have that as $n\to \infty$ 
\begin{align*}
&E\{(y_{ij}-\varepsilon_{ij})^2\}=O(\Psi/n)\quad \mbox{for all $i< j$}; \\
&E\{(y_{ij}-\varepsilon_{ij})(y_{ik}-\varepsilon_{ik})\}=O\{\Psi/n^2+\Upsilon/n\}\\
&E\{(y_{ik}-\varepsilon_{ik})(y_{jk}-\varepsilon_{jk})\}=O\{\Psi/n^2+\Upsilon/n\}  \quad \mbox{for all $i< j< k$}; \ \mbox{ and}\\
&E\{(y_{ij}-\varepsilon_{ij})(y_{kl}-\varepsilon_{kl})\}=O\{\Psi/n^3+\Upsilon/n^2\}\quad \\
&\hspace{4.2cm} \mbox{for all $i< j$ and $ k< l;\ i \neq j\neq k \neq l$}
\end{align*}
under (A-i).
Then, we have that as $m\to \infty$ 
\begin{align}
&\Var(U_n-\widehat{T}_n)=E[\{U_n-(\widehat{T}_n-\Delta)\}^2]=o(K^2) 
\label{A.4}
\end{align}
under (A-i).
Hence, by combining (\ref{A.3}) with (\ref{A.4}), we have that as $m\to \infty$ 
\begin{align*}
\Var(\widehat{T}_n)&=E[\{(T_n-\Delta)-U_n+U_n \}^2]\\
&=\Var( U_{n})+\Var(U_n-\widehat{T}_n)
-2E[\{U_n-(\widehat{T}_n-\Delta)\}U_n]\\
&=K^2\{1+o(1)\}+O(\Omega^{1/2}/n^2)
\end{align*}
under (A-i) from the fact that $|E[\{U_n-(\widehat{T}_n-\Delta)\}U_n]|\le \{\Var(U_n-\widehat{T}_n) \Var(U_n)\}^{1/2}$ by Schwarz's inequality. 
It concludes the result.
\end{proof}
%%%%%%%%%%%%%%%%%%%%%%%%%%%%%%%%%%
\begin{proof}[Proof of Lemma 3.2]
%%%%%%%%%%%%%%%%%%%%%%%%%%%%%%%%%%
Let $r_*=\mbox{rank}(\bSig_1^{1/2}\bSig_*)$. 
When we consider the singular value decomposition of $\bSig_1^{1/2}\bSig_*$, 
it follows that $\bSig_1^{1/2}\bSig_*=\sum_{j=1}^{r_*}\lambda_{*j}\bh_{*j(1)}\bh_{*j(2)}^T$, 
where $\lambda_{*1} \ge\cdots\ge \lambda_{*r_*}(> 0)$ denote singular values of $\bSig_1^{1/2}\bSig_*$, 
and $\bh_{*j(1)}$ (or $\bh_{*j(2)}$) denotes a unit left- (or right-) singular vector corresponding to $\lambda_{*j}\ (j=1,...,r_*)$. 
Then, it holds that 
\begin{align*}
\Upsilon&=\tr(\bSig_1^{1/2}\bSig_* \bSig_2 \bSig_*^T\bSig_1^{1/2})\\
&=\tr \Big\{\Big(\sum_{j=1}^{r_*}\lambda_{*j}\bh_{*j(1)}\bh_{*j(2)}^T\Big)\bSig_2\Big(\sum_{j=1}^{r_*}\lambda_{*j}\bh_{*j(2)}\bh_{*j(1)}^T\Big)\Big\}\\
&=\sum_{j=1}^{r_*}\lambda_{*j}^2\bh_{*j(2)}^T\bSig_2\bh_{*j(2)}\le \lambda_{\max}(\bSig_2)\sum_{j=1}^{r_*}\lambda_{*j}^2= \lambda_{\max}(\bSig_2)\tr(\bSig_*^T\bSig_1 \bSig_*).
\end{align*}
Similarly, we claim that $\tr(\bSig_*^T\bSig_1 \bSig_*)\le  \lambda_{\max}(\bSig_1)\tr(\bSig_*^T\bSig_*)=\lambda_{\max}(\bSig_1)\Delta$, 
so that 
\begin{align}
\Upsilon\le  \lambda_{\max}(\bSig_1)\lambda_{\max}(\bSig_2)\Delta.\label{A.5}
\end{align}
Thus under (A-iii), it holds that $\Upsilon=o(\Delta \Psi^{1/2})$ as $p \to \infty$. 
Then, we claim that $n\Upsilon/\Psi=o(n\Delta/\Psi^{1/2} )$ as $p \to \infty$, 
so that under (A-iii) and (A-v)
\begin{align}
n\Upsilon/\Psi=o(1)\quad \mbox{as $m\to \infty$}.
\label{A.6}
\end{align}
By noting that $\sum_{i=1}^q(\bgamma_{1i}^T\bSig_*\bgamma_{2i} )^2\le \Upsilon$ and $\tr\{(\bSig_* \bSig_*^T)^2\}\le \Upsilon $, 
from Lemma 3.1 and (\ref{A.6}), we have that as $m\to \infty$ 
$$
\Var(\widehat{T}_n)/\delta^2=1+o(1)
$$
under (A-i), (A-iii) and (A-v) from the fact that $\Delta^2/\Psi=o(1)$ as $p \to \infty$ under (A-v).
\end{proof}
%%%%%%%%%%%%%%%%%%%%%%%%%%%%%%%%%%%%
\begin{proof}[Proof of Theorem 3.1]
%%%%%%%%%%%%%%%%%%%%%%%%%%%%%%%%%%%%
From (\ref{A.5}), it holds that $\Upsilon \le \Psi^{1/2}\Delta$, so that $K^2=O(\Psi/n^2+ \Psi^{1/2}\Delta/n)$. 
Then, from Lemma 3.1 and $\Omega^{1/2}\le \Psi$, it holds that as $m\to \infty$  
$$
\Var(T_{n}/\Delta)=O\{\Psi/(n^2\Delta^2)+\Psi^{1/2}/(n\Delta)\}
$$
under (A-i).
Thus, under (A-iv), from Chebyshev's inequality, we can claim the result. 
\end{proof}
%%%%%%%%%%%
We give the following lemmas to prove Theorem 3.2. 
\\[5mm]
%%%%%%%%%%%%%%%%%%%%%%%%
{\it {\bf Lemma A.1.} \ 
It holds that
\begin{align*}
&E(\omega_{ij}^2\omega_{i'j}^2)=O(\Psi^2) \quad \mbox{for all $i,i'\neq j$}; \mbox{ and}\\
&E(\omega_{ij}\omega_{i'j}\omega_{ij'}\omega_{i'j'} )=O(\Omega )\quad \mbox{for all $i\neq i'\neq j \neq j'$}
\end{align*}
under (A-ii).
}
%%%%%%%%%%%%%%%%%%%%%%%%
\begin{proof}
We first consider the first result of Lemma A.1. 
Let $\zeta_{rstu}=  \bgamma_{1r}^T\bgamma_{1t}\bgamma_{2s}^T\bgamma_{2u}$ for all $r, s,t, u$. 
Let $A_1=\sum_{r\neq s}^q\sum_{t\neq u}^q \zeta_{rstu}(\zeta_{rstu}+\zeta_{srtu}+\zeta_{rsut}+\zeta_{srut}) w_{ri}^2w_{si}^2w_{tj}^2w_{uj}^2$ and $A_2=\omega_{ij}^2-A_1$. 
Note that $E(A_1)=B$ and $E(A_2)=0$ under (A-ii).
Here, we claim that $\sum_{r\neq s}^q\sum_{t\neq u}^q  (\zeta_{rstu}^2+\zeta_{srtu}^2+\zeta_{rsut}^2+\zeta_{srut}^2)=O(\Psi)$, so that
$$
\sum_{r\neq s}^q\sum_{t\neq u}^q(|\zeta_{rstu}|+|\zeta_{srtu}|+|\zeta_{rsut}|+|\zeta_{srut}|)^2=O(\Psi).
$$
Then, under (A-ii), we have that 
\begin{align}
E(A_1^2)&\le E\Big\{\Big(\sum_{r\neq s}^q\sum_{t\neq u}^q (|\zeta_{rstu}|+|\zeta_{srtu}|+|\zeta_{rsut}|+|\zeta_{srut}|)^2w_{ri}^2w_{si}^2w_{tj}^2w_{uj}^2\Big)^2\Big\}\notag \\
&=O(\Psi^2). \label{A.7}
\end{align}
For $E(A_2^2)$, it is necessary to consider the terms of $w_{ri}^3w_{r'i}^3w_{r''i}^2\ (r\neq r' \neq r'')$ because it does not hold that $E(w_{ri}^3w_{r'i}^3w_{r''i}^2)=0\ (r\neq r' \neq r'')$ unless $E(w_{ri}^3)=0$ or $E(w_{r'i}^3)=0$. 
Here, under (A-ii), we evaluate that for sufficiently large $C>0$
\begin{align*}
&\Big|E\Big(\sum_{r\neq r' \neq r'' }^q \sum_{t\neq u}^q
\zeta_{rr'tu}\zeta_{rr''tu}\sum_{t'\neq u'}^q \zeta_{rr't'u'}\zeta_{r'r''t'u'}w_{ri}^3w_{r'i}^3w_{r''i}^2 w_{tj}^2w_{uj}^2w_{t'j}^2w_{u'j}^2  \Big)\Big|\\
&\le C \sum_{r\neq r' \neq r'' }^q \sum_{t\neq u}^q
|\zeta_{rr'tu}\zeta_{rr''tu}|\sum_{t'\neq u'}^q |\zeta_{rr't'u'}\zeta_{r'r''t'u'}|  \\ 
&\le C \sum_{r, r', r'' }^q \Big\{ \Big( \sum_{t, u}^q\zeta_{rr'tu}^2\Big) \Big( \sum_{t, u}^q\zeta_{rr''tu}^2\Big)\Big\}^{1/2}
\Big\{ \Big( \sum_{t, u}^q\zeta_{rr'tu}^2\Big) \Big( \sum_{t, u}^q\zeta_{r'r''tu}^2\Big)\Big\}^{1/2} \\ 
&\le C \Big\{ \sum_{r, r', r'' }^q\Big( \sum_{t, u}^q\zeta_{rr'tu}^2\Big) \Big( \sum_{t, u}^q\zeta_{rr''tu}^2\Big)\Big\}^{1/2}
\Big\{\sum_{r, r', r'' }^q \Big( \sum_{t, u}^q\zeta_{rr'tu}^2\Big) \Big( \sum_{t, u}^q\zeta_{r'r''tu}^2\Big)\Big\}^{1/2} \\ 
&\le C \Big( \sum_{r, r' }^q \sum_{t, u}^q\zeta_{rr'tu}^2\Big)\Big( \sum_{r, r''}^q\sum_{t, u}^q\zeta_{rr''tu}^2\Big)^{1/2}
\Big( \sum_{r', r''}^q\sum_{t, u}^q\zeta_{r'r''tu}^2\Big)^{1/2}=O(\Psi^2)
\end{align*}
from the fact that $|E(w_{ri}^3)|\le  \{E(w_{ri}^4)E(w_{ri}^2)\}^{1/2}\le M_r^{1/2}$ for all $r$. 
Similarly, for other terms, we can evaluate the order as $O(\Psi^2)$. 
Hence, we can claim that $E(A_2^2)=O(\Psi^2)$ under (A-ii), so that
$$
E(\omega_{ij}^4)=O\{E(A_1^2)+E(A_2^2)\}=O(\Psi^2)
$$
from (\ref{A.7}). 
By noting that $E(\omega_{ij}^2\omega_{i'j}^2) \le \{E(\omega_{ij}^4) E(\omega_{i'j}^4)\}^{1/2}$, 
we can conclude the first result of Lemma A.1. 

Next, we consider the second result of Lemma A.1. 
From (\ref{A.1}), under (A-ii), we can evaluate that 
$$
E(\omega_{ij}\omega_{i'j}\omega_{ij'}\omega_{i'j'} )=O(\Omega )+O(\Upsilon^2)=O(\Omega )\quad \mbox{for all $i\neq i'\neq j \neq j'$}.
$$
It concludes the second result of Lemma A.1. The proof is completed. 
%%%%%%
\end{proof}
\noindent
%%%%%%%%%%%%%%%%%%%%%%%%
{\it {\bf Lemma A.2.} \ 
It holds that as $m\to \infty$
$$
Var(\widehat{T}_n-V_n)=o(\delta^2)
$$
under (A-i), (A-iii) and (A-v).
}
%%%%%%%%%%%%%%%%%%%%%%%%
\begin{proof}
From (\ref{A.2}), 
we have that under (A-i)
\begin{align*}
&E\{(\omega_{ij}-\varepsilon_{ij})^2\}=O(\Upsilon+\Omega^{1/2})\quad \mbox{for all $i\neq j$}; \\
&E\{(\omega_{ij}-\varepsilon_{ij})(\omega_{ik}-\varepsilon_{ik})\}=O(\Upsilon)  \quad \mbox{for all $i\neq j\neq k$}; \\
&\mbox{and }\ E\{(\omega_{ij}-\varepsilon_{ij})(\omega_{kl}-\varepsilon_{kl})\}=0\quad  \mbox{for all $i\neq j\neq k\neq l$}. 
\end{align*} 
Then, from (\ref{A.6}), we have that as $m\to \infty$ 
\begin{align}
&Var(U_n-V_n)=O(\Upsilon/n+\Omega^{1/2}/n^2)=o(\delta^2) 
\label{A.8}
\end{align}
under (A-i), (A-iii) and (A-v).
By combining (\ref{A.8}) with (\ref{A.4}), from the fact that $\Var(\widehat{T}_n-V_n)=O\{ \Var(\widehat{T}_n-U_n) +\Var(U_n-V_n)\}$, 
we can conclude the result. 
\end{proof}
%%%%%%%%%%%%%%%%%%%%%%%%%%%%%%%%%%%%
\begin{proof}[Proof of Theorem 3.2]
Let 
$
v_{j}=2\{n(n-1)\}^{-1}\sum_{i=1}^{j-1}\omega_{ij}
$
for $j=2,...,n$. 
Note that $\sum_{j=2}^n v_{j}=2\sum_{i< j}^n \omega_{ij}/\{n(n-1)\}=V_n$. 
Also, note that 
$$
\Var\Big(\sum_{j=2}^n v_{j}\Big)=4\sum_{i< j}^n \frac{E(\omega_{ij}^2)}{n^2(n-1)^2}=\frac{2B}{n(n-1)}.
$$ 
Here, we have for $j=3,...,n$, that $E(v_{j}|v_{j-1},...,v_{2})=0$. 
Then, we consider applying the martingale central limit theorem given by McLeish \cite{McLeish:1974}. 
Let $\xi_j=v_{j}/[2B/\{n(n-1)\}]^{1/2}$, $j=2,...,n$. 
Note that $\sum_{j=2}^nE(\xi_j^2)=1$ and $\Var(\sum_{j=2}^n \xi_j)=1$.
Let $I(\cdot)$ denote the indicator function. 
From (\ref{A.1}), under (A-ii) and (A-iii), 
it holds that as $p\to \infty$  
\begin{equation}
B=\Psi+\Delta^2+O(\Omega^{1/2})=\Psi\{1+o(1)\}+\Delta^2.\label{A.9}
\end{equation}
Then, by using Chebyshev's inequality and Schwarz's inequality, from Lemma A.1, under (A-ii) and (A-iii), it holds for Lindeberg's condition that as $m\to \infty$ 
\begin{align}
\sum_{j=2}^{n} E \{  \xi_{j}^2 I ( 
 \xi_{j}^2 \ge \tau  ) \} \le  \sum_{j=2}^{n} \frac{E(\xi_{j}^4)}{\tau}= O\Big( \frac{\Psi^2}
{B^2n } \Big)\to 0
\label{A.10}
\end{align}
for any $\tau>0$. 
Here, from Lemma A.1, (\ref{A.9}) and (\ref{A.10}), under (A-ii) and (A-iii), we evaluate that as $m\to \infty$ 
\begin{align*}
&\sum_{j=2}^{n}E[\{\xi_{j}^2-E(\xi^2)\}^2\le \sum_{j=2}^{n}E(\xi_{j}^4)  \to 0; \ \mbox{ and}\\
&\sum_{2\le i<j \le n} E[\{\xi_{i}^2-E(\xi_{i}^2)\}\{\xi_{j}^2-E(\xi_{j}^2)\}]=
O\Big( \frac{\Psi^2}{B^2n }+\frac{\Omega}{B^2} \Big)\to 0,
\end{align*}
so that 
\begin{align}
\Var\Big(\sum_{j=2}^n \xi_j^2\Big)=E \Big[ \Big\{\sum_{j=2}^n \{\xi_j^2-E(\xi_j^2)\} \Big\}^2 \Big]\to 0.
\label{A.11}
\end{align}
Then, by using the martingale
central limit theorem, from (\ref{A.10}) and (\ref{A.11}), 
under (A-ii) and (A-iii), we obtain that as $m\to \infty$
\begin{align}
\frac{V_n}{\sqrt{2B/\{n(n-1)\}}}=\sum_{j=2}^n \xi_{j}\Rightarrow N(0,1).\label{A.12}
\end{align}
Note that $\delta/[2B/\{n(n-1)\}]^{1/2}\to 1$ as $m\to \infty$ under (A-ii), (A-iii) and (A-v). 
Then, by combining (\ref{A.12}) with Lemmas 3.2 and A.2, we have that as $m\to \infty$
\begin{align}
\frac{\widehat{T}_n-\Delta}{\sqrt{\Var(T_n)}}=\frac{\widehat{T}_n-\Delta}{\delta}+o_P(1)=\frac{V_n}{\sqrt{ 2B/\{n(n-1)\}}}+o_P(1)\Rightarrow N(0,1)
\label{A.13} 
\end{align}
under (A-ii), (A-iii) and (A-v).
It concludes the result. 
\end{proof}
%%%%%%%%%%%%%%%%%%%%%%%%%%%%%%%%%%
\begin{proof}[Proof of Lemma 3.3]
By Lemma 3.1 after replacing $(\bSig_2,\bgamma_{2j},\bSig_*,\Delta)$ with $(\bSig_1,$
$\bgamma_{1j},\bSig_1,\tr(\bSig_1^2))$, 
we can conclude the result when $i=1$. 
When $i=2$, we have the result similarly.
Thus the proof is completed. 
\end{proof}
%%%%%%%%%%%%%%%%%%%%%%%%%%%%%%%%%%%%%%
\begin{proof}[Proof of Corollary 3.1]
By combining Theorem 3.2 with Lemma 3.3, we can conclude the result. 
\end{proof}
%%%%%%%%%%%%%%%%%%%%%%%%%%%%%%%%%%%%%%%%%%%%%%%%%%%%%%%
\begin{proof}[Proofs of Theorem 4.1 and Corollary 4.1]
We first consider the proof of Corollary 4.1. 
From Theorem 3.1, under (A-i) and (A-iv), we have that as $m\to \infty$ 
\begin{equation}
P\Big( \frac{\widehat{T}_n}{ \widehat{\delta}} >z_{\alpha}\Big)
=P\Big( \frac{\widehat{T}_n}{\Delta}>z_{\alpha}\frac{\widehat{\delta}}{\Delta}\Big)
=P\Big(1+o_P(1)>o_P(1)\Big)\to 1 \notag
\end{equation}
from the fact that $\widehat{\delta}/\Delta=o_P(1)$ as $m\to \infty$ under (A-i) and (A-iv). 
It concludes the result of Corollary 4.1. 

Next, we consider the proof of Theorem 4.1.
From Corollary 3.1, under (A-ii), (A-iii) and (A-v), we have that as $m\to \infty$ 
\begin{equation}
P\Big( \frac{\widehat{T}_n}{ \widehat{\delta}} >z_{\alpha}\Big)
=P\Big( \frac{\widehat{T}_n-\Delta}{{\delta}}+o_P(1) >z_{\alpha}-\frac{\Delta}{\delta} \Big)
=\Phi\Big(\frac{\Delta}{\delta}-z_{\alpha} \Big)+o(1) \notag
\end{equation}
from the fact that $\widehat{\delta}/\delta=1+o_P(1)$ as $m\to \infty$ under (A-ii). 
We can conclude the results of size and power when (A-v) is met in Theorem 4.1. 
We note that $\Phi({\Delta}/{\delta}-z_{\alpha} )\to 1$ as $m\to \infty$ under (A-iv), so that we obtain the result of power when (A-iv) is met from Corollary 4.1.
Hence, by considering the convergent subsequence of ${\Delta}/{\delta}$, 
we can conclude the result of power in Theorem 4.1. 
The proofs are completed. 
\end{proof}
%%%%%%%%%%%%%%%%%%%%%%%%%%%%%%%%%%%%%%%%
\begin{proof}[Proof of Proposition 5.1]
We first consider the case when (A-iv) is met. 
From Theorem 3.1 and Lemma 3.3, 
under (A-i) and (A-iv), it holds that as $m\to \infty$
$$
\widehat{\kappa}=\frac{\Psi }{n^2 \Delta^2}\{1+o_P(1)\}=o_P(1).
$$
It concludes the result when (A-iv) is met. 

Next, we consider the case when (A-v) is met. 
From (\ref{A.5}), it holds that $\Upsilon\le  \lambda_{\max}(\bSig_1)\lambda_{\max}(\bSig_2)\Delta\le \Psi^{1/2}\Delta$, 
so that $n\Upsilon/\Psi=O(1)$ as $m\to \infty$ under (A-v). 
Then, from Lemma 3.1 and (\ref{A.1}), under (A-i) and (A-v), we claim that $\Var(\widehat{T}_n)=O(\Psi/n^2)$ as $m\to \infty$. 
Note that $\Delta=O(\Psi^{1/2}/n)$ as $m\to \infty$ under (A-v). 
Thus under (A-i) and (A-v), it holds that $\widehat{T}_n=\Delta+O_P(\Psi^{1/2}/n)=O_P(\Psi^{1/2}/n)$ as $m\to \infty$.
Then, from Lemma 3.3, under (A-i) and (A-v), we have that as $m\to \infty$
$$
\widehat{\kappa}^{-1}=\frac{n^2 \widehat{T}_n^2}{\Psi }\{1+o_P(1)\}=O_P(1).
$$
It concludes the result when (A-v) is met. 
The proof is completed. 
\end{proof}
%%%%%%%%%%%%%%%%%%%%%%%%%%%%%%%%%%%%%%%%
\begin{proof}[Proof of Proposition 5.2]
By combining Lemmas 3.1 and 3.3, we can conclude the result. 
\end{proof}
%%%%%%%%%%%%%%%%%%%%%%%%%%%%%%%%%%
\begin{proof}[Proof of Lemma 5.1]
Let $y_{ij,0}=\widehat{\Delta}_{ij,0}+||\bSig_0||_F^2-\Delta_0$ and 
$\varepsilon_{ij,0}=\varepsilon_{ij}-\bx_{1i}^T\bSig_0\bx_{2i}-\bx_{1j}^T\bSig_0\bx_{2j}+2\tr(\bSig_*\bSig_0^T)$
for all $i < j$. 
From (\ref{A.2}), we write that 
\begin{align*}
y_{ij,0}=&\tr\{(\widehat{\bSig}_{*,ij(1)}-\bSig_*)(\widehat{\bSig}_{*,ij(2)}-\bSig_*)^T \} 
\notag \\
&+\tr(\widehat{\bSig}_{*,ij(1)}\bSig_{*0}^T)+\tr(\widehat{\bSig}_{*,ij(2)}\bSig_{*0}^T)-2\tr(\bSig_*\bSig_{*0}^T) 
; \ \mbox{ and}\notag \\
\varepsilon_{ij,0}=&\omega_{ij}+\eta_{ij}+\eta_{ji}+\psi_{ij}+
\tr(\bx_{1i} \bx_{2i}^T\bSig_{*0}^T)
+\tr(\bx_{1j}\bx_{2j}^T\bSig_{*0}^T)-2\tr(\bSig_*\bSig_{*0}^T) 
\notag
\end{align*}
for all $i<j$. Then, in a way similar to the proof of Lemma 3.1, we can conclude the result. 
\end{proof}
%%%%%%%%%%%%%%%%%%%%%%%%%%%%%%%%%%%%%%
\begin{proof}[Proof of Corollary 5.1]
Let $\Upsilon_0=\tr(\bSig_1\bSig_{*0} \bSig_2 \bSig_{*0}^T)$. 
Similarly to (\ref{A.5}), it holds that 
$$
\Upsilon_0 \le \lambda_{\max}(\bSig_1) \lambda_{\max}(\bSig_2)\Delta_0 \le \Psi^{1/2}\Delta_0.
$$
Then, by noting that $\sum_{i=1}^q(\bgamma_{1i}^T\bSig_{*0}\bgamma_{2i} )^2\le \Upsilon_0 $ and 
$\tr\{(\bSig_{*0} \bSig_*^T)^2\}\le \Upsilon_0$, from Lemma 5.1, we have that as $m\to \infty$ 
$$
\Var(T_{n,0}/\Delta_0)=O\{\Psi/(n^2\Delta_0^2)+\Psi^{1/2}/(n\Delta_0)\}
$$
under (A-i).
Thus, under (A-iv) with $\Delta=\Delta_0$, from Chebyshev's inequality, we can claim the result of Corollary 5.1. 
\end{proof}
%%%%%%%%%%%%%%%%%%%%%%%%%%%%%%%%%%%%%%
\begin{proof}[Proof of Corollary 5.2]
Similarly to the proof of Lemma A.2, under (A-i), (A-iii), (A-v) and (A-v) with $\Delta=\Delta_0$, 
we can claim that $\Var(\widehat{T}_{n,0}-V_{n})=o(\delta^2)$ as $m\to \infty$. 
Thus, similar to (\ref{A.13}), from Lemma 3.3, we can conclude the result.
\end{proof}
%%%%%%%%%%
%%%%%%%%%%%%%
\par
\vspace{9pt}
\noindent {\large\bf Acknowledgment}
\par
\vspace{9pt}
Research of the first author was partially supported by Grant-in-Aid for Young Scientists (B), Japan Society for the Promotion of Science (JSPS), under Contract Number 26800078.
Research of the second author was partially supported by Grants-in-Aid for Scientific Research (B) and 
Challenging Exploratory Research, JSPS, under Contract Numbers 22300094 and 26540010. 
\par
\vspace{9pt}
%%%%%%%%%%%%%%%%%%%%%%%%%%%

%\bibliography{mybibfile}

\end{document}